\documentclass[a4paper,11pt]{article}
\pdfoutput=1
\usepackage{jcappub}
\usepackage{epsfig}
\usepackage{amsmath}
\usepackage{amsfonts}
\usepackage{amssymb}
\usepackage{graphicx}
\usepackage{colordvi}
\usepackage{xfrac}
\usepackage{changepage}

\newcommand{\be}{\begin{equation}}
\newcommand{\ee}{\end{equation}}
\newcommand{\bea}{\begin{eqnarray}}
\newcommand{\eea}{\end{eqnarray}}
\newcommand{\nn}{\nonumber\\}

\newcommand{\miso}{\frac{1}{2}}
\newcommand{\alp}{\alpha'}

\newcommand{\llang}{\langle \! \langle}
\newcommand{\rrang}{\rangle \! \rangle}

\def\lsim{\hbox{ \raise.35ex\rlap{$<$}\lower.6ex\hbox{$\sim$}\ }}
\def\gsim{\hbox{ \raise.35ex\rlap{$>$}\lower.6ex\hbox{$\sim$}\ }}

\def\lsim{\hbox{ \raise.35ex\rlap{$<$}\lower.6ex\hbox{$\sim$}\ }}
\def\gsim{\hbox{ \raise.35ex\rlap{$>$}\lower.6ex\hbox{$\sim$}\ }}

\title{\boldmath The D-material universe}
\vspace{1cm}
\author[a]{Thomas Elghozi}
\author[a,b]{Nick E. Mavromatos}
\author[a]{Mairi Sakellariadou}
\author[c]{Muhammad Furqaan Yusaf}

\affiliation[a]{Theoretical Particle Physics and Cosmology Group, Department of Physics, King's College London, University of London, Strand, London WC2R 2LS, United Kingdom}
\affiliation[b]{Theory Division, Physics Department, CERN, CH-1211 Geneva 23, Switzerland.}
\affiliation[c]{Theoretical Physics and Astrophysics Groups, School of Physics, H. H. Wills Physics Laboratory, University of Bristol, Tyndall Avenue, Bristol, BS8 1TL, United Kingdom}

\emailAdd{thomas.elghozi@kcl.ac.uk}
\emailAdd{nikolaos.mavromatos@kcl.ac.uk}
\emailAdd{mairi.sakellariadou@kcl.ac.uk}
\emailAdd{furqaan.yusaf@bristol.ac.uk}

\date{\today}

\abstract {In a previous publication by some of the authors (N.E.M., M.S. and M.F.Y.), we have argued that the ``D-material universe'', that is a model of a brane world propagating in a higher-dimensional bulk populated by collections of D-particle stringy defects, provides a model for the growth of large-scale structure in the universe via the vector field in its spectrum. The latter corresponds to D-particle recoil velocity excitations as a result of the interactions of the defects with stringy matter and radiation on the brane world. In this article, we first elaborate further on the results of the previous study on the galactic growth era and analyse the circumstances under which the D-particle recoil velocity fluid may ``mimic" dark matter in galaxies. A lensing phenomenology is also presented for some samples of galaxies, which previously were known to provide tension for modified gravity (TeVeS) models. The current model is found in agreement with these lensing data. Then we discuss a cosmic evolution for the D-material universe by analysing the conditions under which the late eras of this universe associated with large-scale structure are connected to early epochs, where inflation takes place. It is shown that inflation is induced by dense populations of D-particles in the early universe, with the r\^ole of the inflaton field played by the condensate of the D-particle recoil-velocity fields under their interaction with relativistic stringy matter, only for sufficiently large brane tensions and low string mass scales compared to the Hubble scale. On the other hand, for large string scales, where the recoil-velocity condensate fields are weak, inflation cannot be driven by the D-particle defects alone. In such cases inflation may be driven by dilaton (or other moduli) fields in the underlying string theory.}

\begin{document}

\begin{flushleft} 
KCL-PH-TH/2015-\textbf{53}, \, LCTS/2015-40 \\
\end{flushleft}

\maketitle
\flushbottom

%%%%%%%%%%%%%%%%%%%%%%%%%%%%%%%%%%%%%%%%%%%%%%%%%%%%%%%%%%
\section{Introduction and summary}

November this year (2015) celebrates a century after Einstein proposed his equations for the dynamics of the gravitational field and the induction of space-time curvature by matter, which established the extremely successful field of General Relativity. Several important predictions from this theory have been verified experimentally in a plethora of experiments or observations, ranging from Earth-based ones to Astrophysical/Cosmological. Moreover, the current year has been declared the international year of light. In this respect we should mention that, over the past two decades, observations on cosmic light from celestial sources at cosmological distances from Earth have revealed important information on the structure and properties of the early universe. Extreme precision measurement of cosmic microwave background radiation fluctuations~\cite{Planck}, as well as observations of baryon acoustic oscillations~\cite{bao} at galactic scales and high-redshift extragalactic type IA supernovae~\cite{sna} lead, upon basing the underlying theory on a Friedmann-Lema\^itre-Robertson-Walker (FLRW) universe that follows the cosmological version of Einstein equations, to important information on the universe's current energy budget, its evolution and properties to as early epochs as inflation.~\cite{Planck_infl}. Nevertheless, there are still many important issues that remain open or not understood at all. One of the most pressing ones, concerns the nature of the dark sector of our universe. Indeed, the above-mentioned observations~\cite{Planck}, lead to the conclusion that only 5\% of the universe energy budget is occupied by the known matter (atoms, particles of the Standard Model and radiation). The rest is conjectured to be of unknown origin and in particular 69\% of the current-epoch energy budget represents an unknown ``dark energy'' component, while the remaining 26\% corresponds to an unknown form of matter, termed ``dark matter''. The latter has been attributed to (yet undiscovered) particles that may exist in extensions of the standard model, such as axions, supersymmetry, string theory, higher-dimensional field theories, etc... Although there are stringent constraints on the dark matter relic abundance from direct and indirect (including collider physics) searches, nevertheless there is still no concrete experimental evidence of the existence of such particles.

The lack of such direct experimental evidence, prompted conjectures that a dark matter component in the universe does not exist but, instead, the assumption that the Newtonian gravitational equations describe the universe at galactic scales should be relaxed: one may have a \emph{MO}dified \emph{N}ewtonian \emph{D}ynamics (MOND) at such scales~\cite{mond}. MOND Theories have been embedded in relativistic modified gravitational field theories, where in additional to graviton fields, one had extra scalar and (constrained) vector modes, the so-called \emph{Te}nsor-\emph{Ve}ctor-\emph{S}calar (TeVeS) theories of gravity~\cite{teves}. The phenomenology of these modified gravity theories is, at present, controversial, in the sense that although agreement is claimed in general with the plethora of galactic data~\cite{famay}, nevertheless at least the simplest models of TeVeS theories proposed initially~\cite{teves} are not compatible with lensing data in some sets of galaxies, including the bullet cluster, in the sense that one needs significant amounts of dark matter to make these models consistent with the data~\cite{ferreras}; this in contradiction to their original motivation as alternatives to dark matter. The situation of course is far from being conclusive, given that there may still still (more complicated) phenomenological models of modified gravity, or improvements in our knowledge of mass profiles and other parameters of such galaxies, that can still allow TeVeS models to be compatible with these data. The cosmology of TeVeS theories has also been developed~\cite{tevescosmo} and some interesting links of the vector fields of such theories with large scale growth of the universe have been proposed~\cite{dodelson}. Of course their phenomenology is not yet as well studied (or their agreement with the entirety of the available cosmological data as good) as the ``conventional'' $\Lambda$CDM (Cosmological Constant ($\Lambda$) Cold Dark Matter) model. One may argue, of course, that this might be due to the fact that the currently used TeVeS models are too simple for that purpose as they do not address the fundamental issue of the origin of the dark energy, which, if achieved, might lead to more complex theories that could fit the data better. It should be noted, however, that there is no microscopic origin of the currently available modified gravity (TeVeS/MOND) models, based on some underlying fundamental physics, and this is in our opinion a major drawback of all such models.

In ref.~\cite{ms} we conjectured that modified gravity models involving fundamental vector fields, but quite different from TeVeS models, may appear as the low-energy limit of certain brane theories of the type proposed in ref.~\cite{dfoam}. According to such theories, a (3~+~1)-dimensional brane universe propagates in a higher dimensional bulk punctured by populations of (effectively point-like) D0-brane (D-particle) defects, which are only weakly interacting among themselves, basically through gravitational interactions, and thus behave more or less as a dark matter cosmic fluid~\cite{mitsou} (termed \emph{D-matter} in \cite{dmatter}). In ref.~\cite{msy} we have put forward a proposal for the r\^ole such populations of D-particles might play in our universe, viewed as a D-brane world, in regards to large scale structure. We have argued that the fluctuations of the recoil velocity of (populations of) D-particle defects, arising from their interaction with stringy matter~\cite{kogan,gravanis,mitsou,dflation}, can provide the seeds for the formation of galaxies, provided their densities are larger than a critical value. Formally this has been demonstrated by representing these recoil fluctuations as mean-field vector excitations of a stringy $\sigma$-model that describes the propagation of strings in cosmological FLRW space-time backgrounds, punctured by populations of fluctuating D-particles. It should be stressed that, since these models are based on string theory, their low-energy effective actions may contain phenomenologically realistic extensions of the standard model, which include conventional dark matter candidates, such as neutralinos, axions, etc... Our point of view in ref.~\cite{msy} is that the D-particle recoil velocity fluid may provide an additional component that play the r\^ole of a mixture (as evidenced from the respective equation of state) of dark-matter-dark-energy component, which was shown to be responsible for large-scale structure growth, in a remote analogy with the r\^ole of the vector fields of TeVeS models~\cite{dodelson}. However, we should stress that our D-brane/D-particle cosmologies are unrelated to, and in fact are very different from, TeVeS models in both their dynamics and spectra.

In the current work, we extend further the study of such models by discussing the r\^ole that the D-particles can play in two distinct eras of the universe evolution that are large structure and galaxy formation, and inflation. For galactic dynamics, we show that their effects can ``mimic" and duplicate the effect of dark matter, under some circumstances that are outlined in detail. Specifically, data from galactic lenses indicate a miss-match between the observed baryonic mass content of the galaxy and the strength of the lensing it produces. This discrepancy is usually accounted for by the inclusion of dark matter gravitational sources, however we are able to show that D-particles can play a similar r\^ole. Their statistically averaged recoil velocities induce a modification to the standard gravitational relation, enhancing its effect on these scales. This mechanism can then compliment and enhance any dark matter component which may be present.

As we shall discuss in this article, we impose constraints on the density of the D-particles on our brane world and fine tune the cosmological constant so that the cosmic concordance model $\Big(\Omega_m, \, \Omega_\Lambda, \, \Omega_k \Big)$ (in a standard notation), which best fits the observations~\cite{Planck,bao,sna}, is satisfied within experimental errors for the galactic era. The spatial flatness $\Omega_k=0$ is guaranteed in our brane world model by construction (which is also consistent with our inflationary scenario), viewing our universe as a spatially flat brane. In general of course, without any restriction, the evolution of the D-particle Universe could be very different from the standard $\Lambda$CDM model. For the current era, of relevance for lensing, the so-considered range of D-particle densities occurs within a small window, which also ensures perturbation growth and large-scale structure but does not imply fine tuning of the model's parameters. Moreover, as we shall also show in this work, the age of the Universe in our model matches the expected one from the $\Lambda$CDM model.

Furthermore, we also study the r\^ole that the condensation of the recoil velocity field can have in providing a (slow-roll) inflation mechanism, consistent with the latest cosmological data such as the Planck survey~\cite{Planck_infl}. As we shall discuss, for the inflationary scenario to be driven by the D-particle-recoil velocity condensates, one needs large condensate fields, which require dense populations of D-particle defects in both the bulk and the brane world in the early universe epochs, as well as low string mass scales compared to the Hubble scale of the pertinent inflation. For weak condensates, such as the ones characterising the galactic era, where the string scale is large compared to the Hubble inflationary scale, inflation \emph{cannot} be driven by the D-particles, although in such scenarios other moduli fields of string theory, such as the dilaton, might do the job, and in fact provide a Starobinsky-type inflation~\cite{dflation}. We also provide a smooth connection, in the sense of a cosmic evolution, between the weak condensates of the recoil velocity field at late epochs of the universe relevant for galaxy and large-scale structure formation, with the strong condensates induced by dense D-particle populations in the early universe that drive inflation. This is what we call the \emph{D-material universe}.

The structure of the paper is as follows: in section~\ref{sec:actions} we review the relevant formalism and give the corresponding effective actions that describe the dynamics of the D-material universe. Given that the requirement of D-particle-driven inflation requires different parameters of the model (such as brane tensions) than those examined in ref.~\cite{msy}, which concentrated only on weak condensates sufficient for lensing data, we revisit the relevant dynamics and outline the relevant conditions that allow the D-particle fluids to ``mimic" via their recoil fluctuations the dark matter component. This is done in section~\ref{sec:lensing}, where by performing a lensing phenomenology in a sample of galaxies we demonstrate consistency of the approach, in the sense that the required densities of D-particles from galaxy to galaxy vary only by at most 10\%-20\%, which is compatible with the existing uncertainties of the lensing data. Finally in section~\ref{sec:inflation} we discuss the r\^ole of D-particles in driving inflation and show that only for dense populations in the early universe, which implies large condensate fields, this is possible. Nevertheless, as already mentioned, the case of weak condensates is not ruled out since in this case inflation might be driven by the dilaton or other moduli fields present in string theory. Finally, conclusions and outlook are given in section~\ref{sec:conclusions}. Some technical aspects and other review material of our approach are described in two appendices.

%%%%%%%%%%%%%%%%%%%%%%%%%%%%%%%%%%%%%%%%%%%%%%%%%%%%%%%%%%
\section{Low-energy string effective actions for the D-material universe} \label{sec:actions}

As discussed in refs.~\cite{msy,dflation}, the following four-dimensional space-time effective action expresses the interaction of stringy matter on a brane world of three longitudinal large dimensions with a medium of recoiling D-particles in the early universe\footnote{Throughout this article, the following conventions are adopted: the metric signature is $(-, +, +, +)$, the Riemann curvature tensor is defined as $R^\alpha _{\;\beta\gamma\delta} = \partial_\delta \,\Gamma^\alpha_{\beta\gamma} + \Gamma^\lambda_{\beta\gamma} \,\Gamma^\alpha_{\lambda\delta} - (\gamma \leftrightarrow \delta)$ and the Ricci tensor as $R_{\mu\nu} = R^\alpha_{\;\mu \nu \alpha}$.}
\begin{align} \label{action1}
	S_{\rm eff~4D} = \int d^4x &\left[ - \frac{1}{4} \,e^{-2 \phi} \,{\mathcal G}_{\mu\nu} \,{\mathcal G}^{\mu\nu} - \frac{T_3}{g_{{\rm s}0}} \,e^{- \phi} \,\sqrt{- \det \left( g + 2\pi\alp F \right)} \left( 1 - \alpha R(g) \right) \right. \nonumber \\
	& - \left. \sqrt{-g} \,\frac{e^{-2 \phi}}{\kappa_0^2} \,{\tilde \Lambda} + \sqrt{-g} \,\frac{e^{-2 \phi}}{\kappa_0^2} \,R(g) + \mathcal{O} \left( \left( \partial \phi \right)^2 \right) \right] + S_{\rm m} ~,
\end{align}
where the second term on the right-hand-side is the standard Dirac-Born-Infeld (DBI) action describing the dynamics of open vector fields on a D-brane world; $S_m$ denotes the matter action, describing the dynamics of matter and radiation particles on the brane world; ${\mathcal G}_{\mu\nu}$ is a flux gauge field, one of the many that in general exist in brane models, and whose importance will become clear later on; $T_3$ is the brane tension, a priori unconstrained; $g =\det (g_{\mu\nu})$ is the determinant of the gravitational field $g_{\mu\nu}$; $\phi$ is the dilaton field, which is assumed constant, $\phi=\phi_0$, for our purposes; and $g_{{\rm s}0}$ is the string coupling for constant dilaton $\phi_0$. In the presence of a dilaton, the full string coupling is defined as $g_{s} = g_{{\rm s}0} \,e^\phi$.

The vector field $A_\mu$ (of mass dimension one, in our conventions) denotes the recoil velocity field excitation during the string-matter/D-particle interactions~\cite{kogan,gravanis} which has a field strength (of mass dimension two) given by
\be \label{fieldstrength}
	F_{\mu\nu} \equiv \partial_\mu A_\nu - \partial_\nu A_\mu ~.
\ee
The background configurations will be discussed later on. We consider the recoil gauge field mainly confined on the brane world and hence its dynamics are described by the well known Born-Infeld square root Lagrangian on the D3-brane world, which includes a resummation over all powers of $\alpha^\prime$~\cite{tseytlin}.

The quantity $\kappa_0^2$ is the four-dimensional bulk-induced gravitational constant defined as
\be \label{bigc}
	\frac{1}{\kappa_0^2} = \frac{V^{(6)}}{g_{{\rm s}0}^2}~ M_{\rm s}^2 ~,
\ee
where $V^{(6)}$ is the compactification volume in units of the Regge slope $\alpha^\prime$ of the string theory describing the excitations on the brane world and
\be \label{defa}
	\alpha = \alp \zeta (2) = \alp \pi^2/6 ~,
\ee
in the (open string/brane) model of ref.~\cite{Cheung}, which we adopt here. The cosmological constant term $\tilde \Lambda$ is induced in general by bulk physics~\cite{dflation} and it is a free parameter in the phenomenological approach we follow here. Note that in certain models, it may even be of anti-de-Sitter type $\tilde \Lambda < 0$.

Before proceeding we should remark for completeness that a basic assumption~\cite{msy} underlying (\ref{action1}) is that any mass contribution of the D-particle defects to the vacuum energy density is considered subleading, compared to the recoil and other terms present in (\ref{action1}). This is because, as discussed in refs.~\cite{dfoam,mitsou}, there are \emph{mixed sign} contributions to the brane vacuum energy as a consequence of bulk D-particle populations at various distances, near (less than the string scale) or further (a few string lengths) away from the D3 brane world, which ``screen'' the D-particle mass contributions. For future use we mention at this stage that the D-particle mass energy density in the Einstein frame is independent of the dilaton $\phi$. To see this, let us start from the corresponding expression in the string frame
\be \label{dmassdensity}
	\rho^{\rm string}_{\rm D-mass~density} = \frac{M_{\rm s} e^{-\phi}}{g_{{\rm s} 0}} \,\frac{\mathcal N^\star_{\rm{D}}}{{\mathcal V}^{(3)}}
\ee
where $\mathcal N^\star_{\rm{D}}/\mathcal V^{(3)}$ is the number three-density of D-particles on the brane world and ${\mathcal V}^{(3)}$ is the proper three-volume. Above, we took into account that, since (\ref{dmassdensity}) refers to a contribution on the brane world effective action, where open strings end, there is a $e^{-\phi}$ factor in front of the corresponding space-time integral, which makes the effective mass of the D-particle $M_{\rm s}/g_{\rm s} = e^{-\phi}\,M_{\rm s}/g_{{\rm s} 0}$, given that the string coupling is $g_{\rm s} = g_{{\rm s} 0} \, e^\phi$. Upon passing into the Einstein frame (designated by an upperscript ``E'' in the appropriate quantities), that is upon rescaling the space-time metric by (in four space-time dimensions)
\be \label{einstein}
	g_{\mu\nu}^{\rm E} = e^{-2\phi} g_{\mu\nu}
\ee
the proper volume scales as ${\mathcal V}^{(3)\, {\rm E}} = e^{-3\phi} \, {\mathcal V}^{(3)}$, and hence the energy density (\ref{dmassdensity}) contribution to the four-dimensional action (\ref{action1}) becomes
\be \label{dmassdensity2}
	\int d^4x \sqrt{-g} \,\rho^{\rm string}_{\rm D-mass~density} (g) = \int d^4x \sqrt{-g^{\rm E}} \,\frac{M_{\rm s}}{g_{{\rm s} 0}} \frac{\mathcal N^\star_{\rm{D}}}{{\mathcal V^{\rm E}}^{(3)}} \equiv \int d^4 x \sqrt{-g^{\rm E}} \,\rho_{\rm D-mass~density}^{{\rm E}}
\ee
that is independent of the dilaton $\phi$. This will be used below, when we consider constraints on the vacuum energy imposed by using the dilaton equations of motion. Such terms escape these constraints. Note that in the Einstein frame, the mass of the D-brane is fixed to $M_{\rm s}/g_{{\rm s}0}$. Moreover, it is in this frame that the standard FLRW form of the universe metric is assumed, implying the scaling $\rho_{\rm D-mass~density}^{{\rm E}} \propto a^{-3}$, where $a$ is the scale factor, on account of the assumption of ``weak- or no-force condition'' among D-particles~\cite{dfoam,mitsou}.\footnote{Numerically, if the contribution (\ref{dmassdensity2}) was unscreened from bulk D-particle effects, it could play the r\^ole of a dark matter energy density component. In such a case, and under the assumption that today there are $n_{\rm D}^{\star \, (0)}$ D-particles per (reduced) Planck three-volume $M_{\rm Pl}^3$ in the brane world, we would have in terms of the critical density today $\Omega_{\rm D~matter} = \frac{M_{\rm Pl}^2}{3 \,H_0^2} \,\frac{M_{\rm s}}{g_{{\rm s}0} \,M_{\rm Pl}} \,n_{\rm D}^{\star \,(0)}$, where $H_0$ is today's value of the Hubble parameter. If this were the dominant dark matter component, then, to avoid overclosure of the universe, we should demand that $\Omega_{\rm D~matter} \lsim \Omega_m^{(0)} $, where $\Omega_m^{(0)}\sim 0.3$ is the current value of the matter contribution in the universe energy budget (in units of the critical density), as measured by astrophysical data~\cite{Planck}. Thus we obtain the following upper bound for today's density of the D-particle populations on the brane world
\be \label{Nstar}
	n_{\rm D}^{\star \, (0)} < 0.9 \,\frac{M_{\rm Pl}}{M_{\rm D}} \,\frac{H_0^2}{M_{\rm Pl}^2} ~,
\ee
where $M_{\rm D}=M_{\rm s}/g_{{\rm s}0}$. Given that $H_0/M_{\rm Pl} \sim 10^{-60}$, and $M_{\rm s} \ge \mathcal O(10)$ TeV, phenomenologically, we observed that this bound is very stringent. However, as mentioned earlier, the presence of bulk D-particles, interacting via stretched open strings with the brane world, as well as the bound D-particles, screen the D-particle mass to an effective one $M_{{\rm D}} \ll M_{\rm s}/g_{{\rm s}0}$. In this case, much higher densities on the brane world are allowed, as can be seen from the above considerations.} Thus, in general, taking into account their scaling as $a^{-3}$, one may absorb such subleading D-particle mass terms contributions (\ref{dmassdensity2}) into the `matter' action $S_m$. Throughout this work we assume that the D-particle recoil velocity effects to the vacuum energy --- which by the way, as we showed in ref.~\cite{msy} and discuss here (cf. eq.~(\ref{u2}) in appendix \ref{sec:bfe}), also scale like $a^{-3}$ after exit from inflation --- dominate until the current era.\footnote{In other scenarios, even those effects may be screened by the bulk D-particle populations that accompany the D3-brane world in its travel through the bulk space, and in this way much higher densities of D-particles bound on our brane universe can be allowed without overclosing the universe~\cite{mitsou}. In fact, in such cases one may use the induced refractive index of the (dense) ``D-foam'' medium at late eras in order to explain certain observed delays of the more energetic photons compared to lower energy ones, from distant active celestial sources such as Gamma Ray Bursts or Active Galactic Nuclei~\cite{refractive}. On the other hand, under the assumption adopted in the current work that the recoil effects are dominant until the current era, one obtains as we shall see (cf. eq.~(\ref{range})) an upper bound for the allowed D-particle densities that is much weaker than the one required to reproduced these delays.} During inflation, as we discussed in \cite{dflation} and shall review below, the assumed high density of D-particles implies a constant density of D-particles on the brane, which contribute crucially to a Starobinsky-like inflation driven by strong condensate fields of the recoil velocities field strength $\llang F_{\mu\nu} \, F^{\mu\nu} \rrang$.

After this necessary digression, we next remark that the four-dimensional DBI action (on the D3-brane world) in (\ref{action1}) can be expanded in derivatives, as appropriate for a low-energy weak-field approximation\footnote{Such derivative expansions are appropriate for weak recoil fields, which is the case characterising the galactic eras of the universe, of interest to us when we consider the r\^ole of the D-particles as providers of structure growth~\cite{msy}. However, as we shall discuss later in the article, dense D-particle populations in the early eras can condense and induce an inflationary era, which is characterised by strong recoil velocity fields. The latter necessitates keeping the Born-Infeld-square-root structure intact.} compared to the string scale $M_{\rm s} = \sfrac{1}{\sqrt{\alp}}$, as follows~\cite{tseytlin}
\be \label{dbi2}
	{\det}_4 \left( g_{\mu\nu} + 2\pi \alp F_{\mu\nu} \right) = {\det }_4 g \left[ 1 + (2 \pi \alp)^2 I_1 - (2 \pi \alp)^4 I_2^2 \right]~,
\ee
where
\be \label{i2a}
	I_1 = \miso \,g^{\mu\lambda} g^{\nu\rho} F_{\mu\nu} F_{\lambda\rho}~, \quad I_2 = - \frac{1}{4} \epsilon^{\mu\nu\lambda\rho} F_{\mu\nu} F_{\lambda\rho} ~.
\ee
In our approach, as discussed in appendix~\ref{sec:bfe}, the ``magnetic field'' dual components for the recoil velocity field strength are subleading, and thus for us $I_2$ will not be considered further. Upon such derivative expansions, the resulting effective action on the D3-brane universe, in the Einstein frame (\ref{einstein}) (denoted by a superscript $E$), becomes
\begin{align} \label{actionSmallCondensate}
	S^{\rm E}_{\rm eff~4D} = \int {\rm d}^4x \sqrt{-g} &\left[ \frac{- T_3 \,e^{3 \phi_0}}{g_{{\rm s}0}} - \frac{\tilde{\Lambda} \,e^{2 \phi}}{\kappa_0^2} - \frac{(2\pi \alpha^\prime)^2 \,T_3 \,e^{3 \phi_0}}{g_{{\rm s}0}} \,\frac{F^2}{4} \,\left( 1 - \alpha \,e^{-2 \phi_0} \,R \right) \right. \\
	& + \left. \left( \frac{\alpha \,T_3 \,e^{\phi_0}}{g_{{\rm s}0}} + \frac{1}{\kappa_0^2} \right) R + \dots \right] + S_{\rm m} ~, \nonumber
\end{align}
where from now on we work with a constant dilaton $\phi = \phi_0$.

As we discussed in ref.~\cite{msy} and review in the appendix, there is an additional constraint which the velocity field satisfies; it basically arises from the standard relativity relation of a four velocity $u^\mu u_\mu = -1 $ where the contraction is with the metric $g_{\mu\nu}$
\be \label{constraint2}
	A_\mu \,A_\nu \,g^{\mu\nu} = - \frac{1}{\alp} ~,
\ee
where the right-hand side arises from dimensional considerations.

For a constant dilaton, which is the case we consider throughout this article, one may redefine the vector field $A_\mu$ so as to have canonical kinetic (Maxwell-type) term, that is
\be \label{renorm}
	A_\mu \to {\widetilde A}_\mu \equiv \sqrt{\frac{(2 \pi \alp)^2 \,T_3 \,e^{3\phi_0}}{g_{{\rm s}0}}} \;A_\mu ~,
\ee
which implies that the constraint (\ref{constraint2}) becomes
\be \label{constraint3}
	\widetilde A_\mu \,\widetilde A_\nu \,g^{\mu\nu} = - \frac{4 \pi^2 \alp \,T_3 \,e^{3\phi_0}}{g_{{\rm s}0}} ~.
\ee
Thus, from now on we shall be dealing with the action
\begin{align} \label{action2b}
	S^{\rm E}_{\rm eff~4D} = \int {\rm d}^4x \sqrt{-g} &\left[ -\frac{T_3 \,e^{3 \phi_0}}{g_{{\rm s}0}} - \frac{\tilde{\Lambda} \,e^{2 \phi_0}}{\kappa_0^2} - \frac{\widetilde F^2}{4} \left( 1 - \alpha \,e^{-2 \phi_0} \,R \right) + \left( \frac{\alpha \,T_3 \,e^{\phi_0}}{g_{{\rm s}0}} + \frac{1}{\kappa_0^2} \right)\, R \right. \nn
	& + \left. \lambda \left( \tilde A_\mu \tilde A^\mu + \frac{4 \pi^2 \alp \,T_3 \,e^{3 \phi_0}}{g_{{\rm s}0}} \right) \dots \right] + S_{\rm m} ~,
\end{align}
where $\widetilde F$ is the Maxwell field strength for the field $\widetilde A_\mu$ given in eq.~(\ref{renorm}) and $\lambda$ is a Lagrange multiplier implementing the constraint (\ref{constraint3}). Notice that the first two terms on the right-hand-side of eq.~(\ref{action2b}) play the r\^ole of a cosmological constant
\be \label{cc}
	\Lambda_0 \equiv \frac{ T_3 \,e^{3 \phi_0}}{g_{{\rm s}0}} + \frac{\tilde{\Lambda} \,e^{2 \phi_0}}{\kappa_0^2} ~.
\ee
A detailed study of the equations of motion and background solutions for the recoil vector field have been discussed in ref.~\cite{msy} and will only be briefly reviewed below, as we shall need them to make our estimates on the D-particle recoil-velocity fluctuation effects on the galactic dynamics. The background field configurations we shall use for our purposes in this article are reviewed in appendix~\ref{sec:bfe}, along with their basic properties and order of magnitude estimates.

The graviton equation of motion obtained from eq.~(\ref{action2b}), on assuming to first approximation that terms involving $\rho_{\rm vac} \ll \widetilde F^2$ or derivatives of $\widetilde F^2$ can be neglected,\footnote{This is because spatial derivatives yield terms proportional to spatial derivatives of the metric, that is proportional to the Newtonian acceleration $\zeta^\prime$ of a D-particle in the gravitational field of the galaxy, while temporal derivatives yield terms proportional to the Hubble parameter today $H_0$, which are again suppressed compared with the terms that are kept.} is given by
\begin{align} \label{lenseineqn}
	\left( R^\mu_{~\nu} - \frac{\delta^\mu_{~\nu}}{2} R \right) \left[ \frac{\alpha \,T_3 \,e^{\phi_0}}{g_{{\rm s}0}} + \frac{1}{\kappa_0^2} + \frac{\alpha e^{-2\phi_0} \,\widetilde F^2}{4} \right] - \frac{1}{2} \left( 1 - \alpha e^{-2 \phi_0} R \right) \widetilde F^{\mu \lambda} \widetilde F_{\nu \lambda} & \nn
	+ \,\frac{1}{8} \widetilde F^2 \,\delta^\mu_{~\nu} + \lambda \,\widetilde A^\mu \widetilde A_\nu - \lambda \,\frac{\delta^\mu_{~\nu}}{2} \left( \widetilde A_\alpha \widetilde A^\alpha + \frac{1}{\alpha'} {\cal J} \right) + \frac{1}{2} \delta^\mu_{\,\nu} \,\Lambda_0 &= \frac{1}{2} \,T^\mu_{~\nu}~,
\end{align}
where $T_{\mu\nu} = -2\frac{1}{\sqrt{-g}} \,\frac{\delta S_m}{\delta g^{\mu\nu}} $ is the matter stress tensor and where we defined
\be \label{defJ}
	{\mathcal J} \equiv \frac{(2 \pi \alp)^2 \,T_3 \,e^{3\phi_0}}{g_{{\rm s}0}} ~.
\ee
For future use we note that, by contracting the vector field equation of motion obtained from eq.~(\ref{action2b})
\be \label{veom}
	\left[ \widetilde F_{\nu \mu} \left( 1 - \alpha \,e^{-2 \phi_0} R \right) \right]^{;\nu} + 2 \lambda(x) \,\widetilde A_\mu = 0
\ee
(where the semicolon denotes covariant derivative) with $A^\mu$, and then applying the constraint (\ref{constraint3}), we obtain the following form for the (background value) of the Lagrange multiplier field
\be \label{lm}
	\langle \lambda (x) \rangle = \frac{e^{-3\phi_0} \,g_{{\rm s}0}}{8\pi^2 \alp \,|T_3|} \,\widetilde A^{\mu} \left[ \widetilde F_{\nu \mu} \left( 1 - \alpha e^{-2\phi_0} R \right) \right]^{;\nu}~,
\ee
which we shall make use in the next section when we estimate the recoil effects of the D-particles on the universe dynamics during the galactic era. An important point to note, which was the subject of discussion in ref.~\cite{msy} and which we only mention here for completeness, is that the constraint term in eq.~(\ref{action2b}) is vital in coupling the recoil vector field perturbation to the density perturbations through the $\langle \lambda (x) \rangle \widetilde A_\mu \widetilde A_\nu$ term in the respective stress-energy tensor obtained from eq.~(\ref{action2b}).

Last but not least, before we embark on an estimation of the D-particle effects on the galactic dynamics, we would like to comment on the physical importance of the constraints imposed by the dilaton equation, which should be taken into account despite the fact that the dilaton is considered to be constant in our analysis. The dilaton equation of motion, obtained by varying the effective action (\ref{action2b}) with respect to the dilaton field $\phi$ and setting it to a constant value $\phi_0$ at the end of the variation, reads
\begin{align} \label{dileq}
	\frac{3 \,T_3 e^{3 \phi_0}}{g_{{\rm s}0}} + \frac{2 \,\tilde{\Lambda} \,e^{2\phi_0}}{\kappa_0^2} - \alpha \left[ \frac{T_3 e^{\phi_0}}{g_{{\rm s}0}} - \frac{e^{-2\phi_0} \,\widetilde F^2}{2} \right] R + \ldots = 0~,
\end{align}
where the $\ldots$ on the left-hand-side of this equation denote subleading terms (involving derivatives on $\widetilde F$ of the form $- \frac{3}{2} \alpha e^{-2 \phi} \nabla^2 (F_{\alpha \beta} F^{\alpha \beta})$ or proportional to the Lagrange multiplier $\lambda$ which satisfies eq.~(\ref{lm})) which are ignored. Taking into account~\cite{msy} that for galactic scales the terms $\alpha \,\widetilde F \widetilde F \,R \ll \alpha \,\frac{T_3 e^{\phi_0}}{g_{{\rm s}0}} \,R \ll 3 \,\frac{T_3 e^{3 \phi_0}}{g_{{\rm s}0}}$, eq.~(\ref{dileq}) can be well approximated by
\be \label{t3ltilde}
	\frac{T_3}{g_{{\rm s}0}} \,e^{3\phi_0} \simeq - \frac{2}{3} \,\frac{\tilde \Lambda \,e^{2\phi_0}}{\kappa_0^2}~,
\ee
which in turn implies that the cosmological constant on the brane world $\Lambda_0$, defined in eq.~(\ref{cc}) with positive tension $T_3 > 0$ (as required for stability), is negative
\be \label{negLambda}
	\Lambda_0 \simeq -\frac{1}{2} \frac{T_3}{g_{{\rm s}0}} \,e^{3\phi_0} \,< \,0 ~.
\ee
This can be remedied by assuming that such (anti-de-Sitter type) terms cancel against \emph{dilaton independent contributions} to the brane vacuum energy, coming from appropriate combinations of the mass terms of D-particles bound to the brane world~\cite{dflation} as in eq.~(\ref{dmassdensity2}), and bulk gauge flux fields inducing condensates of the form appearing in the action (\ref{action1}) --- which as we shall argue in section~\ref{sec:inflation} play an important r\^ole for inflation. In this way, during the galactic era, only a \emph{small positive cosmological constant} term survives, which plays no significant r\^ole on the galactic scale lensing phenomenology, in accordance to observations. This assumption will be understood in what follows.

%%%%%%%%%%%%%%%%%%%%%%%%%%%%%%%%%%%%%%%%%%%%%%%%%%%%%%%%%%
\section{Lensing phenomenology of the D-material universe \label{sec:lensing}}

To discuss the phenomenology of our D-particle universe using galactic lensing data, we use the action (\ref{action2b}). We shall use the local form of the recoil vector field eq.~(\ref{ai_gen1}), averaged at the end over populations of D-particles in the neighbourhood of a galaxy. As mentioned previously, we also assume that at the local galactic level, any contribution of the four-dimensional brane world vacuum energy $\Lambda_0$ is small (or cancelled appropriately), of the order of the observed cosmological constant today, so that it may be safely neglected to a first approximation when considering the dynamics at galactic scales, as relevant for lensing.

%----------------------------------------------------------------------------------%
\subsection{The equations of motion}

Let us now give an order of magnitude estimate of the various terms appearing in the graviton equation (\ref{lenseineqn}), setting $\Lambda_0$ to zero, before we proceed with the detailed lensing phenomenology. This will be useful in yielding a qualitative understanding on the order of magnitude of the quantity $|\beta|$, defined in eq.~(\ref{stat}), needed for the D-particle defects to play the r\^ole of dark matter candidates and providers of large scale growth structures~\cite{msy}.

To this end, we first notice that, upon using classical backgrounds, which are discussed in detail in ref.~\cite{msy} and reviewed briefly in the appendix, the penultimate term on the left-hand-side of eq.~(\ref{lenseineqn}) will vanish identically, due to the constraint (\ref{constraint3}) that such backgrounds satisfy. In addition, on account of eq.~(\ref{lm}), the term on the stress tensor proportional to the Lagrange multiplier $\lambda$ yields terms proportional to derivatives of the field strength, which are suppressed compared to the remaining contributions from the vector field (however the reader should bear in mind the aforementioned important r\^ole of this term in coupling the perturbations of the vector field to the density perturbations~\cite{msy}, thus leading to the growth of structure).

Moreover, as we discussed above, any spatial (``magnetic type'') components of the field strength are subleading compared to the ``electric'' type ones $F_{0i}$, which, on account of (\ref{foicorrect1}), implies
\begin{align} \label{foi}
	\widetilde F^{t\rho} \widetilde F_{t\rho} = - \,a^2(t) \,e^{\zeta(r) - \nu(r)} \,\delta_{jk} u^j u^k \,\frac{\mathcal J}{\alpha'^2} \left( \frac{a^2(t_c)}{a^2(t)} - 2 H(t) \,t_c \right)^2 ~,
\end{align}
where the local metric that contracts the velocities is given by eq.~(\ref{localmetric}). For a lensing galaxy at redshift $z$, we apply eq.~(\ref{foicorrect}) for today's observational time ($t=t_0$, $a(t_0)=1$)~\cite{timered}
\begin{align} \label{foic}
	\widetilde F^{t\rho} \widetilde F_{t\rho} = - \,e^{\zeta(r) - \nu(r)} \,\delta_{jk} u^j u^k \,\frac{\mathcal J}{\alpha'^2} \,{\mathcal H}(z)^2~,
\end{align}
with ${\mathcal H}(z) = \left[ \frac{1 - 3(1 + z)^2}{(1 + z)^2 \,(1 + (1 + z)^2)} \right]$.

At this point we can take the statistical average of the velocities over populations of D-particles in the neighborhood of galaxies, as given in eqs.~(\ref{stat}), (\ref{u2}), which yields
\begin{align}
	\llang \widetilde F^{t\rho} \widetilde F_{t\rho} \rrang = - {\mathcal J} \,\frac{3 \,\sigma_0(t)^2}{\alpha'^2} \,e^{\zeta(r)-\nu(r)} \,{\mathcal H}(z)^2~,
\end{align}
where the index $i$ spans Cartesian coordinates. We also have (in Cartesian coordinates)
\begin{eqnarray}
	\llang \widetilde F_\mu^{\,\,\rho} \widetilde F_{\nu\rho} \rrang = {\mathcal J} \,\frac{\sigma_0^2}{\alp^2} \left[ \begin{array}{cccc}
		-3 e^{\zeta-\nu} & \dots & \dots & \dots \\
		\dots & -e^{\zeta-\nu} & \dots & \dots \\
		\dots & \dots & -e^{\zeta-\nu} & \dots \\
		\dots & \dots & \dots & -e^{\zeta-\nu}
	\end{array} \right] \,{\mathcal H}(z)^2~,
\end{eqnarray}
where the $\dots$ denote subleading terms of order either $x^i \frac{\zeta^\prime}{r} \,{\mathcal H}(z)$, or $x^i x^j \frac{\zeta'^2}{r^2}\,{\mathcal H}(z)^2$, which are ignored to a first approximation.

Einstein's equations for the lensing system are best described in spherical polar coordinates $(t,r,\theta,\phi)$ to which the above quantities easily transform to. When investigating the modified Einstein's equations, the two components we will use to find the differential equation system that defines $\zeta(r)$ and $\nu(r)$ will be the $tt$ and $\theta\theta$ components. In fact, the symmetry of the system we are analysing allows us to set $\theta=\frac{\pi}{2}$. Thus, the $\theta\theta$ component of $\widetilde F^{\mu\rho} \widetilde F_{\nu\rho}$ will be given by
\begin{align}
	\llang \widetilde F^{\theta\rho} \widetilde F_{\theta\rho} \rrang = - {\mathcal J} \,\frac{\sigma_0^2}{\alp^2} \,e^{\zeta(r)-\nu(r)} \,{\mathcal H}(z)^2 ~, \qquad {\rm with} \quad T^{\theta}_{~\theta} = T^{z}_{~z} \quad {\rm when} \quad \theta = \frac{\pi}{2} ~.
\end{align}
Hence, the components we will use in eq.~(\ref{lenseineqn}) will assume the form
\begin{align} \label{fs}
	\llang \widetilde F^{\alpha\beta} \widetilde F_{\alpha\beta} \rrang &\simeq - 6 \,{\mathcal J} \,\frac{\sigma_0^2}{\alp^2} \,e^{\zeta(r)-\nu(r)} \,{\mathcal H}(z)^2~, \nonumber \\
	\llang \widetilde F^{t \rho} \widetilde F_{t \rho} \rrang &\simeq - 3 \,{\mathcal J} \,\frac{\sigma_0^2}{\alp^2} \,e^{\zeta(r)-\nu(r)} \,{\mathcal H}(z)^2 ~, \nonumber \\
	\llang \widetilde F^{\theta\rho} \widetilde F_{\theta\rho} \rrang &\simeq - {\mathcal J} \,\frac{\sigma_0^2}{\alp^2} \,e^{\zeta(r)-\nu(r)} \,{\mathcal H}(z)^2 ~,
\end{align}
with $\sigma_0^2 (t_0) \simeq |\beta|$ (cf. eq.~(\ref{stat})).

Moreover, non-minimal terms in eq.~(\ref{lenseineqn}) of the form $\widetilde F^2 R $ are also suppressed and ignored in our leading order estimates below. Taking the above considerations into account and concentrating on the $tt$ component of the graviton equation (\ref{lenseineqn}) after taking statistical averages as in eq.~(\ref{stat}) and using eq.~(\ref{fs}), we approximate it as follows (the observation time is set to today $t=t_0$ with $a(t_0)=1$)
\begin{align} \label{lens2}
	\left[ \frac{1}{\kappa_0^2} + \frac{\alpha \,T_3 \,e^{\phi_0}}{g_{{\rm s}0}} + \frac{\alpha \,e^{- 2 \phi_0} \,\llang \widetilde F^2 \rrang}{4} \right] \left( R^t_{~t} - \frac{1}{2} \,R \right) + \frac{1}{8} \,\llang \widetilde F^2 \rrang - \frac{1}{2} \,\llang \widetilde F^{t\rho} \widetilde F_{t\rho} \rrang + \dots &= \frac{1}{2} \,T^t_{~t} \nonumber \\
	\Rightarrow \quad \left[ \frac{1}{\kappa_0^2} + {\cal J} \,\frac{e^{-2 \phi_0}}{24 \,\alpha'} \left( 1 - 6 \pi^2 \,|\beta| \,{\mathcal H}(z)^2 \right) \right] \left( R^t_{~t} - \frac{1}{2} \,R \right) + \frac{3}{4 \,\alp ^2} \,{\mathcal J} \,|\beta| \,{\mathcal H}(z)^2 + \dots &\simeq \frac{1}{2} \,T^t_{~t}~,
\end{align}
where the $\dots $ denote subleading terms, such as $\alpha e^{-2 \phi_0} \,\llang \widetilde F^{t\rho} \,\widetilde F_{t\rho} \rrang \,R$ and terms containing derivatives on $\widetilde F$. For ease of presentation above we also took into account that for the lensing data $e^{\zeta(r)-\nu(r)} =\mathcal{O}(1)$, however for the full numerical calculation presented in Table~\ref{betaT3hern} these terms were computed explicitly.

To model the lensing systems we shall be looking at, we take the energy momentum tensor to describe an ideal pressureless fluid, thus
\begin{align} \label{fluid}
	T^t_{~t} = -\rho(r) ~, \qquad T^i_{~j} = 0~.
\end{align}
We hence observe from the right-hand-side of eq.~(\ref{lens2}) that the recoil-velocity-field contribution to the $tt$ component of the stress tensor has the right sign to be interpreted as a positive energy density contribution.

The quantity
\be \label{egc}
	\frac{1}{\kappa^2_{\rm eff}} \equiv \frac{1}{\kappa_0^2} + {\cal J} \,\frac{e^{-2 \phi_0}}{24 \,\alpha'} \left( 1 - 6 \pi^2 \,|\beta| \,{\mathcal H}(z)^2 \right)
\ee
plays the r\^ole of an effective inverse gravitational constant, which thus depend on the statistical variance of the recoil field $|\beta|$. For the lensing analysis, $|\beta|$ is determined from eq.~(\ref{betagala}) and $M_{\rm s}$ (the string mass) can take any value such that $M_{\rm s}\lesssim~10^{18}$ GeV; however the assumption of non observation of large extra dimensions in current particle accelerators (including the run II of LHC) means that $M_{\rm s}\gtrsim 10^4$ GeV.

For concreteness, from now on we set the constant dilaton value to zero $\phi_0=0$. We also note at this stage that, in the analysis of ref.~\cite{msy}, the brane tension was taken to satisfy
\be \label{setting}
	\frac{(2\pi \alp)^2 \,T_3}{g_{{\rm s}0}} = 1 ~.
\ee
In such a case, $\mathcal J = 1$ in eq.~(\ref{defJ}). However, as we shall see in section~\ref{sec:inflation}, one cannot obtain consistent inflation for brane tensions for which eq.~(\ref{setting}) is adopted. On the contrary, for the case of consistent inflation (that is when we have large-fields, compared to the Planck mass scale) in which the D-material universe evolution connects smoothly the galactic-structure era to the inflationary era, one needs $\mathcal J \gg 1$. This will be the case of interest to us in the present study. Again for concreteness, we consider for the remainder of the paper that the brane tension and the parameter $\kappa_0$ (which is phenomenological in our construction) satisfy
\be \label{choice}
	\kappa_0^{-2} \sim \frac{1}{2} \,M_{\rm Pl}^2 \sim \frac{\mathcal{J}}{24} \,M_{\rm s}^2 ~.
\ee
The identification of the parameter $\kappa_0^{-2} $ with half of the (square of the) four-dimensional reduced Planck mass, $M_{\rm Pl}^2 = \left( 16 \pi G \right)^{-1}$, was chosen so as, on account of eq.~(\ref{egc}), to have
\be \label{keff}
	\kappa_{\rm eff}^{-2} \simeq M_{\rm Pl}^2 ~,
\ee
given that the term proportional to $\beta$ on the right-hand-side of eq.~(\ref{egc}) is relatively suppressed for $|\beta| \ll 1$ (which as we shall see is the case). This is desirable from the point of view of not having significant variations of the gravitational constant, due to the unobserved (so far) violations of the weak equivalence principle. The reader should also notice that the choice eq.~(\ref{choice}) necessitates low string mass scales, $M_{\rm s} \ll M_{\rm Pl}$, as assumed in ref.~\cite{msy}, if one requires $\mathcal J \gg 1$ in order to satisfy the criterion for a smooth connection of this era with inflation.\footnote{If the latter is relaxed, one may consider more general cases, in which the string scale can be as large as $M_{\rm Pl}$~\cite{msy} (in such a case the first two terms in eq.~(\ref{egc}) contribute more or less equally, but the last one is still suppressed for small $|\beta|$).}

Using eq.~(\ref{fluid}), demanding the recoil-vector-field contributions to the stress tensor to be \emph{at most} of the same order of magnitude as the mass terms, and considering typical values of the mass density $\rho (r)$ for lenses to be of order $\rho (r) \sim 10^{-119} \,M_{\rm Pl}^4$ (cf. Table~\ref{comp2} below), we obtain from eq.~(\ref{lens2}) the following \emph{upper bound} on the parameter $|\beta|$
\be \label{upper}
	|\beta| \le \frac{4}{3} \,{\mathcal J}^{-1} \,10^{-119} \,(\mathcal H)^{-2} \left( \frac{M_{\rm Pl}}{M_{\rm s}} \right)^4 \sim 10^{-120} \,\,(\mathcal H)^{-2} \left( \frac{M_{\rm Pl}}{M_{\rm s}} \right)^2~,
\ee
on account of eq.~(\ref{choice}). For e.g. $M_{\rm s} \simeq 10^{4}$ GeV (used as a concrete case in ref.~\cite{msy}) and taking into account that $\mathcal H~\in~(-1, \,-1/3)$, i.e. of ${\mathcal O}(1)$, for a redshift range of interest $z \in (0, 2)$, we observe from eq.~(\ref{upper}) that $|\beta| \le 10^{-92}$.

However, as discussed in ref.~\cite{msy}, there is a \emph{minimum} $|\beta|$, i.e. a minimum density of D-particles, that guarantees the existence of a growing model. The reader should bear in mind that the normalisation of $|\beta|$ in ref.~\cite{msy} was different from the one used in the present study, since in ref.~\cite{msy} we had absorbed in the definition of the recoil velocity a dimensionless factor $t_{\rm c}/\sqrt{\alp}$ where $t_{\rm c}$ is the time of contact of the string matter with the D-particle defect. More precisely, the relation between the recoil velocity $v^{{\rm MSY}}$ used in ref.~\cite{msy} and the dimensionless recoil velocity $u^i$ used here is
\be
	v^{{\rm MSY} \,i} = u^i \,\frac{t_c}{\sqrt{\alp}} ~. \nonumber
\ee
For the galactic era, of interest to us in this section (as well as in ref.~\cite{msy}), $t_{\rm c} \sim t_0 \sim H_0^{-1}$ and thus the relation between the two $|\beta|$ parameters, defined through eq.~(\ref{stat}) for the respective recoil velocities, is given by
\be \label{betas}
	|\beta| \sim \alp H_0^2 \,|\beta^{{\rm MSY}}| ~,
\ee
with again the notation $|\beta^{{\rm MSY}}|$ referring to definitions used in ref.~\cite{msy}.

It was shown in ref.~\cite{msy} that, for $M_{\rm s} \sim 10^4$~GeV, $g_{{\rm s}0} \sim 0.1$, growth of structure due to the recoil velocity field is possible for
\be
	|\beta^{\rm MSY}| \ge 10^{-3} ~, \nonumber
\ee
a value which is largely insensitive to the value of $M_{\rm s} \in (10^4 ; \,10^{18} )$~GeV. This implies (for the $|\beta|$ used here)
\be \label{minbeta}
	|\beta| \ge 10^{-3} \left( \frac{H_0}{M_{\rm s}} \right)^2 \,\sim \,10^{-123} \,\left( \frac{M_{\rm Pl}}{M_{\rm s}} \right)^2 \sim 10^{-95} ~,
\ee
taking into account that $H_0 \sim 10^{-60}~M_{\rm Pl}$.

The analysis in ref.~\cite{msy} assumed a brane tension satisfying eq.~(\ref{setting}). As we shall discuss in the next section, inflation can only be driven by large D-particle recoil-velocity condensates which occur for relatively large brane tensions $T_3$ compared to those satisfying eq.~(\ref{setting}). It would be therefore essential to repeat the growth-of-structure analysis of ref.~\cite{msy} for such large values of $T_3$.

Indeed, relaxing this condition and considering a wider range of values for the tension does not affect the value that $\beta^{\rm MSY}$ needs to take in order to ensure the growth of structure. This is because the dominant terms in the the equation governing the growth of vector perturbations in eq.~\cite{msy} have a simple relationship with the tension $T_3$, such that it appears only as a scaling term. Thus, the vector field associated with the D-particle recoil velocity excitation always enters a growing mode for $\beta^{\rm MSY} \gtrsim 10^{-3}$, and $T_3$ simply scales the result. As a consequence, by appropriately scaling the initial size of the vector perturbations in the early universe, any value of $T_3$ can be made compatible with the growth of structure.

Using eqs.~(\ref{betagala}), (\ref{cmbmom}) in our semi-microscopic model for estimating $|\beta|$, we obtain the allowed range of $|\beta|$ and densities of D-particles. Considering $M_{\rm s} \sim 10^4$ GeV) we get
\be \label{range}
	10^{-95} \le |\beta| \le 10^{-92} \qquad \Rightarrow \quad 10^{-60} \,\widetilde \xi_0^{-2} \,\le \,\frac{n^{(0)}_D}{n^{(0)}_\gamma} \,\le \,10^{-28} \,\widetilde \xi_0^{-2}~, \qquad \widetilde \xi_0 < 1~,
\ee
which serves as an indicative order of magnitude for the required densities so that the D-matter recoil-velocity fluid in this stringy universe can ``mimic'' dark matter in galaxies, in the sense that its contribution to the energy density is of the same order as the mass density of a galaxy. Upon comparing (\ref{range}) with the upper bounds on the number density of D-particles (\ref{Nstar}), obtained in the case where the screening of their mass effects by the D-particles neighboring the D3-brane world did not occur, the alert reader can appreciate the significant increase in the allowed densities in our case, without overclosing the universe.

In this latter respect some remarks are in order at this point. Although the rest mass contributions of D-particles have been assumed to be neutralised to a large extent by repulsive contributions from populations of D-particles in the neighbourhood of the brane world~\cite{msy}, it is worth mentioning that if the above upper bounds on the present-era density of D-particles of (rest) mass $M_{\rm s}/g_{{\rm s}0}$ are satisfied, then the corresponding contributions to the brane's vacuum energy (as seen by a comoving observer, and assuming the D-particles bound on the brane and comoving with its expansion) would be $\rho^{\rm D}_{\rm mass} \sim n_{\rm D}^{(0)} \,M_{\rm s}/g_{{\rm s}0}$. Now, assuming that $n_\gamma^{(0)} = 10^9 n_b^{(0)}$, where $n^{(0)}_{\rm b}$ is the number density of baryons in the universe, and estimating the baryon density by considering a common proton mass $m_p$ for all species of order 1 GeV, we obtain $\rho^{\rm D}_{\rm mass} \sim 10^{-17} \,\rho^{(0)}_{\rm b} \,M_{\rm s} / (g_{{\rm s}0} \,m_{\rm p}) $, with $\rho_{\rm b}^{(0)}$ the energy density of the baryons in the present universe. For masses of $M_{\rm s}/g_{{\rm s}0} \sim 10^5 $~GeV, we obtain that $\rho^{\rm D}_{\rm mass} \sim 10^{-12} \,\rho_{\rm b}$, which means that the overclosing of the universe is in fact not a problem at all.

%----------------------------------------------------------------------------------%
\subsection{The lensing system}

After the above generic estimates, we now proceed with the detailed lensing phenomenology. There are two spherical mass profile which we use for the lensing analysis. First there is the Hernquist mass profile, which is used to model the baryonic mass profiles of the galaxies we are looking at. It is described by
\begin{align} \label{hern}
	M_{\rm H}(\hat{r}) = \frac{M\hat{r}^2}{(\hat{r}+r_h)^2}~.
\end{align}
This definition uses the standard Schwarzschild radius parameter, $\hat{r}$, which is related to the radius parameter which appears in the metric system defined in eq.~(\ref{localmetric}) through $\hat{r} = e^{\zeta(r)/2}r$. There is also a parameter, $r_h$, which defines a scale for the core of the mass distribution. This scale is derived from the observable half mass radius $R_e$. $M$ denotes the total mass of the galaxy.

Second we use the Navarro-Frenk-White (NFW) profile~\cite{NFW}, which is usually used to model the total dark matter and luminous matter contributions to a galaxy. It is described by
\begin{align} \label{nfw}
	M_{\rm NFW}(\hat{r}) = \frac{M}{\Gamma}\left[\ln\left(1+\frac{\mathcal{C}\hat{r}}{r_{\rm vir}}\right)-\frac{\mathcal{C}\hat{r}}{r_{\rm vir}+\hat{r}}\right]~,
\end{align}
where $\Gamma = \ln \left( 1 + \mathcal{C} \right) - \frac{\mathcal{C}}{1 + \mathcal{C}}$, $\mathcal{C}$ is a concentration parameter, usually taken to be around $10$ based on computer simulations, and $r_{\rm vir}$ is the virial radius, related once again to the observable half mass radius, $R_e$.

The above mass profiles are used in the equations describing the deflection of light in our system. The deflection of light in our metric (\ref{localmetric}) is given by
\begin{align}
	\Delta \varphi = 2 \int \limits_{r_0}^\infty \frac{1}{r} \left( e^{\zeta(r)-\nu(r)} \,\frac{r_0^2}{b^2} - 1 \right)^{-1/2}~{\rm d}r - \pi~,
\end{align}
where $r_0$ is the point of closest approach for the light ray and $b$ is the observable impact parameter of the light ray. They are related to each other through
\begin{align}
	b^2 = e^{\zeta(r_0)-\nu(r_0)} \,r_0^2~.
\end{align}
Note that the mass profiles implicitly appear in the above equations through the dependence of $\zeta(r)$ and $\nu(r)$ on the density profile $\rho(r)$, which is derived from the mass profiles. Finally the lensing system is described by the thin lens equation, that is
\begin{align}
	\hat{\beta} = \theta - \Delta \varphi(\theta, M, b) \frac{D_{{\rm ds}}}{D_{\rm s}}~, \label{lensingeq}
\end{align}
where $\hat{\beta}$ is the unknown true angular position of the source galaxy, $\theta$ is the observable angular position of the source, $D_{{\rm ds}}$ is the angular distance from the source to the lens and $D_{\rm s}$ is the angular distance to the source. There are two unknowns in the above equation, the deflection angle $\Delta\varphi$ and $\hat{\beta}$, thus two images of the source are needed and the data from both are combined to constrain the true values of these parameters. Note that here we use a concordance cosmological model $(\Omega_{\rm m}, \Omega_\Lambda, \Omega_k)=(0.3, 0.7, 0)$, since departures from it lead only to insignificant changes in our lensing analysis.\footnote{As we have seen above, the density of D-particles can be constrained appropriately so that the $\Omega_m$ parameter lies within its best-fit value today. The cosmological ``constant'' contribution can also be made small as we discussed previously, below eq.~(\ref{negLambda}). The flatness is of course guaranteed by our brane-world construction.} The lensing equation (\ref{lensingeq}) is applied independently to the multiple images of the background source, and solving it we obtain the actual position $\hat \beta$ of the source and the mass $M$ of the lens.

To estimate the amount of dark matter in the system, the lensing mass is compared to the stellar mass content of the galaxies. The stellar mass is calculated assuming both a Salpeter~\cite{sal} Initial Mass Function (IMF) and a Chabrier~\cite{chab03} IMF. The IMF is defined as the distribution of stellar masses at birth, and for our purposes it is relevant as it dominates the conversion of light into mass. It is usually assumed to be a universal function, although there is some discussion about the exact form of the IMF. For this reason our analysis presents two choices of IMF, a classical Salpeter function, which consists of a single power law, therefore along the lines of a claimed excess of low-mass stars; and a Chabrier IMF, which truncates the power law with a lognormal distribution at the low mass end, resulting in systematic lower values of the mass to light ratio. The stellar mass estimates for the two cases were based on the results presented in ref.~\cite{ig}. The stellar mass is measured out to some apperture radius, and the lensing total mass is truncated to this radius when comparing the two values.

Comparing the lensing mass and the luminous mass of the galaxies allows us to estimate the dark matter content of the galaxies. We can then alter the value of the key parameter in the D-particle model, that is $T_3\beta$, to give the best fit value of $T_3\beta$ for no dark matter to be required in these systems. Note that the purpose of this analysis is not to show the lensing can be explained in the absence of dark matter, as dark matter candidates come naturally with the string model we are working with. However, the results for $T_3\beta$ shown in Table~\ref{betaT3hern} below represent the upper bound on the value of $T_3\beta$ in these systems.

We examine a selection of lensing galaxies from the CfA-Arizona Space Telescope Survey (CASTLES)~\cite{rus03} database. Given our 1-D lensing analysis, we are restricted to only analysing those lenses with 2 images only, as systems with quad images require accounting of the ellipticity of the lens. We are also restricted to looking at those galaxies for which there was high quality data for the luminous mass content. We thus get the list of 11 galaxies for which we present results below. Note that Q0957+561A and Q0957+561B are both a special case for which one galaxy was lensing two separate sources simultaneously. The results for the Hernquist profile are presented in Table~\ref{betaT3hern} and the results for the NFW profile are given in Table~\ref{betaT3nfw}.

Note that in the results, using the Salpeter IMF leads to a negative estimate for the dark matter content of two galaxies, namely BRI0952-012 for the Hernquist and NFW profile, and Q0142-100 for the NFW profile. This is a result of the form of the IMF, which tends to systematically overestimate the contribution of low mass stars when calculating the stellar mass of galaxies. We include both IMF's for completeness and also to show the weak dependence of our results on the specific IMF chosen, but the Chabrier case is widely considered to be the most evidentially robust for precisely this reason~\cite{ig}.

\begin{table}[ht]
\begin{center}
\begin{tabular}{@{}lccccc@{}}
\hline
			&			& \multicolumn{2}{c}{Salpeter IMF}				& \multicolumn{2}{c}{Chabrier IMF}				\\
			&			& ~\% DM~	& $T_3\beta$ for no DM				& ~\% DM~		& $T_3\beta$ for no DM 			\\
Lens			&$z_{\rm l}$		& ~in GR~	& ($\times10^{-121} M_{\rm Pl}^4$)		& ~in GR~		& ($\times10^{-121} M_{\rm Pl}^4$)	\\
\hline
Q0142-100		& $0.49$		& $0.4$	&$7.2\times10^{-3}$ 				& $47.9$		& $8.1\times10^{-1}$			\\
HS0812+123		& $0.39$		& $37.8$ 	& $6.6\times10^{-1}$				& $67.6$		& $1.2$					\\	
BRI0952-012		& $0.63$		& $-6.3$ 	& $-$							& $48.6$		& $7.3	$					\\
LBQS1009-025	& $0.88$		& $64.7$ 	& $1.2$ 						& $81.7$		& $1.5$					\\
B1030+071		& $0.60$		& $59.7$ 	& $1.3$ 						& $78.5$		& $1.7$					\\
HE1104-181		& $0.73$		& $63.2$ 	& $1.1$ 						& $81.9$		& $1.5$					\\
B1152+200		& $0.44$		& $25.1$ 	& $6.3\times10^{-1}$				& $61.0$		& $1.5$					\\
SBS1520+530	& $0.71$		& $41.1$ 	& $8.3\times10^{-1}$				& $67.5$		& $1.4$					\\
B1600+434		& $0.42$		& $61.4$ 	& $1.2$ 						& $78.9$		& $1.6$					\\
HE2149-275		& $0.60$		& $60.7$ 	& $1.1$ 						& $79.7$		& $1.5$					\\
Q0957+561A	& $0.36$		& $76.7$ 	& $1.6$ 						& $86.9$		& $1.8$					\\
Q0957+561B	& $0.36$		& $77.4$ 	& $1.6$ 						& $87.3$		& $1.9$					\\
\hline
\end{tabular}
\caption{The best fit values of $T_3\beta$ to get near zero dark matter for a galaxy using the Hernquist mass profile. Here $M_{\rm s} = 15$ TeV and $z_{\rm l}$ is the redshift of the lensing galaxy. The dark matter requirements in standard GR gravity when comparing the lensing mass to the stellar mass is also given. For details about the different IMFs used here and their meaning, we refer the reader to ref.~\cite{msy}.}
\label{betaT3hern}
\end{center}
\end{table}

\begin{table}[ht]
\begin{center}
\begin{tabular}{@{}lccccc@{}}
\hline
			&			& \multicolumn{2}{c}{Salpeter IMF}				& \multicolumn{2}{c}{Chabrier IMF}				\\
			&			& ~\% DM~	& $T_3\beta$ for no DM				& ~\% DM~	& $T_3\beta$ for no DM				\\
Lens			&$z_{\rm l}$		& ~in GR~	& ($\times10^{-121} M_{\rm Pl}^4$)		& ~in GR~	& ($\times10^{-121} M_{\rm Pl}^4$)		\\
\hline
Q0142-100		& $0.49$		& $-2.9$	& $-$ 							& $46.2$	& $7.8\times 10^{-1}$				\\
HS0812+123		& $0.39$		& $30.2$ 	& $5.3\times 10^{-1}$ 				& $63.6$ 	& $1.1$						\\
BRI0952-012		& $0.63$		& $-3.0$ 	& $-$							& $50.3$	& $7.5$						\\
LBQS1009-025~~	& $0.88$		& $66.9$ 	& $1.2$ 						& $82.8$ 	& $1.5$						\\
B1030+071		& $0.60$		& $61.3$ 	& $1.3$ 						& $79.4$ 	& $1.7$						\\
HE1104-181		& $0.73$		& $54.5$ 	& $9.9\times 10^{-1}$ 				& $77.6$ 	& $1.4$						\\
B1152+200		& $0.44$		& $23.2$ 	& $5.8\times 10^{-1}$ 				& $60.0$ 	& $1.5$						\\
SBS1520+530	& $0.71$		& $43.8$ 	& $8.8\times 10^{-1}$ 				& $69.0$ 	& $1.4$						\\
B1600+434		& $0.42$		& $64.2$ 	& $1.3$ 						& $78.9$ 	& $1.6$						\\
HE2149-275		& $0.60$		& $58.8$ 	& $1.1$ 						& $79.7$ 	& $1.4$						\\
Q0957+561A	& $0.36$		& $74.9$ 	& $1.6$ 						& $86.9$ 	& $1.8$						\\
Q0957+561B	& $0.36$		& $74.3$ 	& $1.6$ 						& $87.3$ 	& $1.8$						\\
\hline
\end{tabular}
\caption{As Table \ref{betaT3hern}, now with the NFW profile.}
\label{betaT3nfw}
\end{center}
\end{table}

Thus, given the allowed range of values for $\beta$ in eq.~(\ref{range}) for $M_{\rm s}=15~{\rm TeV}$, we would expect $T_3$ to be approximately in the range $10^{-58} \,M_{\rm Pl}^4 \lesssim T_3 \lesssim 10^{-26} \,M_{\rm Pl}^4$. This result is actually largely insensitive to the value of the string mass scale $M_{\rm s}$. The corresponding range of $\mathcal J$ (cf. (\ref{defJ})), for $\phi_0=0$, $M_{\rm s} = 15 \,{\rm TeV}$ and phenomenologically relevant values of the string coupling $8\pi^2/g_{{\rm s}0} = \mathcal{O}(10^3)$ is then $10^{12} \lesssim \mathcal J \lesssim 10^{44}$.

%----------------------------------------------------------------------------------%
\subsection{Numerical estimates of the modified contributions}

We shall now examine some numerical results for different contributions of the graviton equation of motion, to allow a more intuitive understanding of the different contributions arising from our model.

\begin{table}[ht]
\begin{center}
\begin{tabular}{@{}c|c@{}}
	$\dfrac{1}{2 \kappa_0^2} + \dfrac{\alpha T_3 e^{\phi_0}}{2 g_{{\rm s}0}} \quad (M_{\rm Pl}^2)$ & $\dfrac{\alpha e^{-2\phi_0} \llang \widetilde F^{\alpha\beta} \widetilde F_{\alpha\beta} \rrang}{4} \quad (M_{\rm Pl}^{-2})$ \\
	\hline
	$4.0\times 10^{-2}$ & $-5.8\,T_3\beta\times 10^{31}$
\end{tabular}
\caption{Comparison of the values for the different contributions to $\kappa^{-2}_{\rm eff}$, defined in eq.~(\ref{egc}). Note that the units in the second column are $M_{\rm Pl}^{-2}$ since $T_3$ itself will come with units of $M_{\rm Pl}^{4}$}
\label{comp1}
\end{center}
\end{table}

Table~\ref{comp1} shows the values of the different contributions to $\kappa^{-2}_{\rm eff}$. The second term will be of the same order as $1/8\pi G$ when $T_3\beta = 10^{-33} \,M_{\rm Pl}^4$. Thus, in order to ensure that there is a negligible contribution from our model to the value of the effective gravitational constant, we get an upper limit on $T_3\beta$ of $10^{-33} \,M_{\rm Pl}^4$. However, more stringent upper limits on this value come from other considerations, as shown in eq.~(\ref{range}).

\begin{table}[ht]
\begin{center}
\begin{tabular}{c|c|c}
	$\left( R^\mu_{~\nu} - \frac{1}{2} \,g^\mu_{~\nu} R \right) \frac{1}{\kappa_0^2} \quad (M_{\rm Pl}^4)$ & $\rho(r) \quad (M_{\rm Pl}^4)$ & $\frac{1}{8} \,\llang \widetilde F^2 \rrang - \frac{1}{2} \,\llang \widetilde F^{t\rho} \widetilde F_{t\rho} \rrang$ \\
	\hline
	$-4.9\times 10^{-120}$ & $1.6\times 10^{-119}$ & 5.8 \,$T_3\beta \times 10^3$
\end{tabular}
\caption{Comparison of the values for the different contributions to the $tt$ component of the metric equations eqs.~(\ref{lens2}), (\ref{fluid}). These contributions are evaluated for the lens HS0818+123, at a distance of $15~{\rm kpc}$. Note that the final column is given in dimensionless form since $T_3$ will come with units of $M_{\rm Pl}^{4}$.}
\label{comp2}
\end{center}
\end{table} 

In Table~\ref{comp2} we show the values of the standard GR terms and the modifications coming from our model. The following can be seen: looking at the last item in the table we can see that this will be of the same order as the GR contribution only when $T_3\beta \approx 1\times 10^{-123}$. Note that the above analysis can only provide a rough understanding of the relative sizes of the contributions coming from different parts of the model; a more accurate numerical analysis presented in Table~\ref{betaT3hern} does not make the same simplifications as the ones that have been taken here, and this accounts for the difference in the calculated value of $T_3\beta$ and the one estimated above.

%%%%%%%%%%%%%%%%%%%%%%%%%%%%%%%%%%%%%%%%%%%%%%%%%%%%%%%%%%
\section{Inflation induced by D-particles \label{sec:inflation}}

Another important aspect of D-particles outlined in ref.~\cite{dflation} is the fact that they may induce inflation, through condensation of their recoil velocity field. There are two physically relevant cases, which as we shall discuss below depend crucially on the size of the string scale involved. One pertains to large condensate fields, which may arise in the case of very dense populations in the early universe, but which dilute today. This case, which we shall study first, is appropriate for low string mass scales $M_{\rm s}$ compared to the Hubble scale. The second case pertains to weak condensates, which are associated with string scales large compared to the Hubble inflationary scale and will be studied later on. Only the strong condensate case leads to consistent inflation, compatible with the Planck data~\cite{Planck}, as we shall demonstrate below.

%----------------------------------------------------------------------------------%
\subsection{Formalism}

Before going to the specifics, it is useful to first introduce the reader to the pertinent formalism. For the inflationary metric, the background recoil-vector field assumes the form (\ref{ai_gen1}) but with the term $t_{\rm c}$ being dominant over the term $(t \,a(t_{\rm c})/a(t))^2$ as a result of the exponential expansion of the universe and the fact that $t_{\rm c}$ is of order of a few string time scales $t_{\rm c} \sim \sqrt{ \alp} $. The relevant background vector field is
\be \label{ai_infl}
	A_i (t) \simeq \frac{1}{\alp} \,g_{ij} (t) \,u^j \left( \frac{a(t_{\rm c})^2}{a(t)^2} \,t - t_c \right) \simeq - \frac{1}{\alp} \,g_{ij} (t) \,u^j \,t_{\rm c} \sim - \frac{1}{\sqrt{\alp}} \,g_{ij} (t) \,u^j ~, \qquad t > t_c \sim \sqrt{\alp} ~,
\ee
and the metric $g_{\mu\nu}(t)$ is just the homogeneous and isotropic de-Sitter FLRW metric (\ref{flrwmetric}) with $a(t) = a_0 \,e^{H_I t}$ and $H_I$ the Hubble scale during inflation, which is assumed constant and of the following order of magnitude $H_I \sim 10^{-5} \,~M_{\rm Pl} \gg H_0$~\cite{Planck}. Thus the recoil vector field is assumed homogenoeus and isotropic in agreement with standard cosmology. This is a feature consistent with the assumption of dense populations of D-particles in the early universe. As discussed previously, we can covariantise the background, using eq.~(\ref{covariant}), which satisfies the constraint (\ref{constraint}). The ``electric'' field strength corresponding to this case is given by eq.~(\ref{ftct}) in appendix~\ref{sec:bfe}. Due to the time dependence only of the background there are no ``magnetic field'' components, $F_{ij}=0$.

As we shall demonstrate below, inflation can be induced in case there are very dense populations of D-particles in the early universe, leading to large condensates of the respective velocity fields. In such a case, the full structure of the Born-Infeld action (\ref{action1}) needs to be kept. As in the perturbative case, we shall also adopt a mean field approach in which we shall consider appropriate distribution functions over the D-particle recoil velocities. We consider distributions of the stochastic Gaussian type (\ref{stat}), which preserve the rotational symmetry and isotropy and homogeneity of space-time (and are thus consistent with the cosmological principle). For strong condensates, the full Born-Infeld effective action (\ref{action1}) will be used to describe inflation. Using eq.~(\ref{dbi2}) we do observe that in the Minkowski space-time, there is an upper bound on the allowed value of the condensate of the ``electric field'' $\llang F_{\mu\nu} \,F^{\mu\nu} \rrang = -\llang \vec E^2 a^2(t)\rrang $, with $F^{0i} \equiv \vec E$, otherwise the integrant of the square root becomes imaginary. This is the ``well-known'' problem of the ``maximal electric field'' in the Minkowskian Born-Infeld theory. However, as we shall discuss below, when we consider the dynamics of inflation, we shall work necessarily in a finite temperature formalism, that is a Euclidean time, in order to account for the (observer dependent) de Sitter temperature characterising the inflationary scenario. In this case, the Euclidean Born-Infled action does not have a bounded ``electric'' field. Analytic continuation back to Minkowski space-time can be performed at the end of the computations. 

From eq.~(\ref{fieldstrength}), one can define a dimensionless covariant condensate in the FLRW space-time background described by the FLRW metric $g_{ij}$ (\ref{flrwmetric})
\begin{align} \label{fst}
	{\mathcal C}(t) &\equiv \left( 2 \pi \alp \right)^2 \llang F_{\mu\nu} \,F^{\mu\nu} \rrang = (2 \pi \alpha^\prime)^2 \;2 \;\llang F_{0i} \,F_{0j} \,g^{00} \,g^{ij} \rrang = - 8 \pi^2 \alp^2 \;\llang F_{0i} \,F_{0j} \,\frac{1}{a^2} \,\delta^{ij} \rrang \nn
	&= - 32 \pi^2 \,\left( \frac{H(t)}{M_{\rm s}} \right)^2 \llang u_i^{\rm phys} u_j^{\rm phys} \,\delta^{ij} \rrang = -96 \pi^2 \,\left( \frac{H(t)}{M_{\rm s}} \right)^2 \,\sigma_0^2 ~,
\end{align}
where $u_i^{\rm phys} = a u_i$ is the D-particle recoil velocity in the comoving frame.

During inflation (approximately de Sitter space-time background), there is a temperature associated with that frame, the so-called Hawking-Gibbons temperature of a de Sitter space-time~\cite{gh}
\begin{equation} \label{HT}
	T = \frac{H}{2 \pi} ~,
\end{equation}
associated with the observer dependent horizon of the de Sitter space-time.

Depending on the relative magnitude of this temperature, we may have a relativistic or non-relativistic ``thermal'' motion of the D-particle ensemble. If $m_{\rm D} = M_{\rm s}/g_{{\rm s}0}$ is the mass of the D-particle, set by the string scale $M_{\rm s}=1/\sqrt{\alp}$, then, for the case where $H \simeq H_{\rm I} \gg m_{\rm D}$, that is when the string scale $M_{\rm s} \ll H_{\rm I} \simeq 10^{14}$~GeV, we have $T/m_{\rm D} \gg 1$ and the thermal motion of D-particles may be considered relativistic. In such a case, a Boltzman distribution of the form
\be \label{boltz}
	\int d^3p \,f(p) \sim \int d^3p \,e^{-p/T} = 4 \pi \int dp \,p^2 \,e^{-p/T}~, \quad p \equiv |\vec p| \gg m_D~,
\ee
may be assumed without loss of generality.

The physical velocity $u_i^{\rm phys} = p_i^{\rm phys} / m_{\rm D} = g_{{\rm s}0} \,p_i / M_{\rm s}$ is now assumed to undergo this thermal distribution, because it is only in the physical (cosmological observer's) frame that such a (observer dependent) temperature can be defined, as explained above. Thus the variance $\llang u_i^{\rm phys} u_j^{\rm phys} \rrang $ (since $\llang u_i ^{\rm phys} \rrang = 0$) can then be computed on the basis of eq.~(\ref{boltz}) to be of order
\begin{align} \label{variance}
	\llang u_i^{\rm phys} u_j ^{\rm phys} \delta^{ij} \rrang &= \frac{1}{m_{\rm D}^2} \,\llang p_i^{\rm phys} p_j ^{\rm phys} \delta^{ij} \rrang = \frac{g_{{\rm s}0}^2}{M_{\rm s}^2} \frac{\int dp \,p^4 \,e^{-p/T}}{\int dp \,p^2 \,e^{-p/T}} \nn
	&\sim \,12 \,g_{{\rm s}0}^2 \left( \frac{T}{M_{\rm s}} \right)^2 \sim \,12 \,g_{{\rm s}0}^2 \left( \frac{H_{\rm I}}{2 \pi M_{\rm s}} \right)^2 ~, \qquad |p| \ll T ~,
\end{align}
during inflation, and that it is approximately constant, decreasing with $H_{\rm I}$.

In the nonrelativistic case, that is when the string scale is much higher than the inflationary scale, one has instead a Maxwell distribution (Gaussian in the velocities) with
\begin{align}
	\int d^3p \,f(p) &\sim \;4 \pi \int dp \,p^3 \,e^{-p^2 / \,T m_{\rm D}}~, \nn
	\llang u_i^{\rm phys} u_j ^{\rm phys} \delta^{ij} \rrang &= \frac{g_{{\rm s}0}^2}{M_{\rm s}^2} \frac{\int dp \,p^5 \,e^{-p^2 / \,T m_{\rm D}}}{\int dp \,p^3 \,e^{-p^2 / \,T m_{\rm D}}} \sim \,2 \,g_{{\rm s}0} \left( \frac{T}{M_{\rm s}} \right) \sim \,2 \,g_{{\rm s}0} \left( \frac{H_I}{2 \pi M_{\rm s}} \right) \ll 1~. \label{variance2}
\end{align}

It is important to notice that the correct treatment of a thermal distribution of recoil velocities requires a Euclideanised space-time, stemming from the replacement of the time coordinate with a Wick rotated one, $x^0 \rightarrow i \tau$, which is then identified with the inverse temperature. In this sense the (dimensionless) condensate of the field strengths (\ref{fst}) then assumes the following form in an order of magnitude estimate
\begin{align} \label{cond}
	{\mathcal C}^{\mathcal E}(t) \sim 96 \,(H_{\rm I}/M_{\rm s})^4 \,g_{{\rm s}0}^2 ~\qquad {\rm for } \quad M_{\rm s} \ll H_{\rm I} &\quad \mbox{(relativistic)} \nn
	\sim 32 \pi (H_{\rm I}/M_{\rm s})^3 \,g_{{\rm s}0} ~\qquad {\rm for } \quad M_{\rm s} \gg H_{\rm I} &\quad \mbox{(non relativistic)},
\end{align}
and notice that this Euclideanised condensate (indicated with the superscript $\mathcal E$) is \emph{positive definite}, so large values (much larger than one) are allowed, which would otherwise have been excluded on the basis of the reality of the argument of the square root of the Minkowskian Born-Infeld Lagrangian. This Euclidean path integral was adopted by Hawking in his treatment of the thermal properties of the black-hole horizon~\cite{hawking}, and by Gibbons and Hawking~\cite{gh} when discussed the de Sitter temperature, which is of interest to us here.

%----------------------------------------------------------------------------------%
\subsection{The fate of D-particle induced inflationary scenarios for small condensates \label{sec:smallcondinfl}}

We next proceed to consider the possibility of D-particle-recoil-induced inflation in the case where the condensate is much smaller than one (weak field), which pertains either to the case when $M_{\rm s} \gg H_{\rm I}$ in eq.~(\ref{cond}) or to the case of arbitrary string scales but with the brane tension satisfying eq.~(\ref{setting}) --- cf. discussion around eq.~(\ref{small}) in the next subsection. Weak condensates characterise the galaxy growth era~\cite{msy}. As we shall demonstrate below, such situations cannot lead to inflation driven by the recoil velocity. Nevertheless, in such a case, other moduli fields in string theory, such as the dilaton (for large negative values), can drive a Starobinsky-type inflation, as discussed in ref.~\cite{dflation} and reviewed briefly in appendix~\ref{sec:dilaton}.

The dynamics of small condensates is described by an appropriate weak-field expansion of the Born-Infeld square root action of (\ref{action1}). For our purposes, it suffices to keep terms up to and including terms quadratic order in the recoil field strength $F^2$, leading to the effective action (\ref{actionSmallCondensate}). The reader should also recall that we have fixed the brane tension to (\ref{setting}) for convenience, which results in a canonical Maxwell kinetic term for the recoil vector field. The latter satisfies the constraint (\ref{constraint}). We also set the dilaton field to zero $\phi_0=0$, since our primary purpose here is to examine the possibility of a D-particle-recoil-driven inflation.

Using the condensation field~\cite{dflation}
\be \label{fcond}
	\llang F^2 \rrang = - 24 \,\sigma_0^2 \,\dot{a}^2 \,\alp = - 24 \,\sigma_0^2 \,H_I^2 \,M_{\rm s}^2 \equiv {\mathcal C}_{\rm M4}~,
\ee
(where ${\mathcal C}_{\rm M4} \equiv \sfrac{1}{(2 \pi \alp)^2} \,{\mathcal C}(t)$ has dimension [mass]$^4$, ${\mathcal C}(t)$ being defined in eq.~(\ref{fst})) and ignoring any matter during the inflationary era, the effective action we shall make use from now on reads
\begin{align} \label{pertaction}
	S_{\rm eff~4D} = \int {\rm d}^4x \,\sqrt{-g} \left[ - \frac{T_3}{g_{{\rm s}0}} - \frac{\tilde{\Lambda}}{\kappa_0^2} - \frac{1}{4} \,{\mathcal C}_{\rm M4} + \left( \frac{1}{\kappa_0^2} + \frac{\alpha T_3}{g_{{\rm s}0}} + \frac{\alpha}{4} \,{\mathcal C}_{\rm M4} \right) R \,\right] ~.
\end{align}
We define then the following constants
\begin{align}
	\frac{1}{\kappa_{\rm{eff}}^2} = \frac{1}{\kappa_0^2} + \frac{\alpha \,T_3}{g_{{\rm s}0}} ~, \qquad \Lambda_0 = \frac{T_3}{g_{{\rm s}0}} + \frac{\tilde{\Lambda}}{\kappa_0^2} \simeq \frac{1}{3} \,\frac{\tilde{\Lambda}}{\kappa_0^2} = f_\kappa \,\frac{\tilde{\Lambda}}{\kappa_{\rm eff}^2} < 0
\end{align}
where for the cosmological constant we used eqs.~(\ref{cc}) and (\ref{t3ltilde}) and defined $f_\kappa \equiv \frac{1}{3} \,\frac{\kappa_{\rm eff}^2}{\kappa_0^2}$. We also define the scalar field
\be \label{sfield}
	\sigma \equiv \frac{1}{4} \,\alpha \,\kappa_{\rm{eff}}^2 \,{\mathcal C}_{\rm M4}
\ee
and we obtain
\begin{align}
	S_{\rm eff~4D} &= \int {\rm d}^4x \,\sqrt{-g} \;\frac{1}{\kappa_{\rm{eff}}^2} \left[ - f_\kappa \,\tilde{\Lambda} - \frac{\sigma}{\alpha} + \left( 1 + \sigma \right) R \,\right] ~.
\end{align}
Since we are concentrating here on the case of small fields, namely $\sigma \ll 1$, this yields
\begin{align}
	\sigma = \frac{1}{4} \;\frac{\pi^2}{6} \frac{1}{M_{\rm s}^2} \,\kappa_{\rm{eff}}^2 \left[ 24 \,\sigma_0^2 \,H_{\rm I}^2 \,M_{\rm s}^2 \right] \simeq \pi^2 \,\sigma_0^2 \left( \frac{H_{\rm I}}{M_{\rm Pl}} \right)^2 &\ll 1~,
\end{align}
where we used $\alpha = \frac{\pi^2}{6} \frac{1}{M_{\rm s}^2}$ and $\frac{1}{\kappa_{\rm{eff}}^2} \simeq M_{\rm Pl}^2$. Recalling $H_I \simeq 10^{-5} \,M_{\rm Pl}$, one has
\begin{align} \label{smallf}
	\sigma_0^2 &\ll \frac{10^{10}}{\pi^2} \simeq 10^9 ~.
\end{align}
Changing the metric $g_{\mu \nu} \rightarrow g^{\rm E}_{\mu \nu} \equiv \left( 1 + \sigma \right) g_{\mu \nu}$ and defining
\begin{align}
	\varphi &\equiv \sqrt{\frac{3}{2}} \;\ln \left( 1 + \sigma \right) \label{DefVarphi} \\
	\partial_\mu \varphi &= \sqrt{\frac{3}{2}} \;\frac{\partial_\mu \sigma}{1 + \sigma} \nonumber
\end{align}
leads to the following
\begin{align}
	R &= \left( 1 + \sigma \right) \left[ R^{\rm E} + \left( \partial_\rho \varphi \,\partial^\rho \varphi \right) \right] \\
	\sqrt{-g} &= \frac{1}{(1 + \sigma)^2} \,\sqrt{-g^{\rm E}} ~.
\end{align}
We hence get the action
\begin{align}
	S^{\rm E}_{\rm eff~4D} &= \int {\rm d}^4x \,\sqrt{-g^{\rm E}} \;\frac{1}{\kappa_{\rm{eff}}^2} \left[ R^{\rm E} + \left( \partial \varphi \right)^2 - V(\varphi) \,\right]~, \nn
	\mathrm{with} \quad V(\varphi) &\equiv \widetilde {\cal D} + \frac{e^{- \sqrt{\frac{2}{3}} \,\varphi}}{\alpha} + \left( f_\kappa \,\tilde{\Lambda} - \frac{1}{\alpha} \right) e^{- \sqrt{\frac{8}{3}} \,\varphi} ~,
\end{align}
where we added a constant of dimension [mass]$^2$ $\widetilde {\cal D} \equiv \kappa_{\rm{eff}}^2 \,{\cal D}$ term corresponding to a flux field condensate (cf. eq.~(\ref{fluxcond}) in the next subsection) as well as other dilaton-independent terms such as the rest mass contributions of a population of D-particles on the brane world to the vacuum energy density~\cite{dflation} (cf. eq.~(\ref{dmassdensity2}), as well as eq.~(\ref{einsteinendens}) in appendix~\ref{sec:dilaton}). We note once again that, as in the large condensate case, it is this field that drives inflation but the fluctuations of the recoil-velocity $\varphi$ inflaton field are what leads to exit from it.

Assuming $\sigma \ll1 $ then $\varphi \simeq \sqrt{1.5} \,\sigma \ll 1$ and thus the field $\varphi$ is also small. Assuming $\varphi \sim 0$ and then Taylor expanding the exponentials, one obtains that the effective (inflationary) potential for small $\varphi \ll 1$ reads
\begin{align} \label{potential}
	V(\varphi) &\simeq \,\left[ \widetilde {\cal D} + \frac{1}{\alpha} \left( 1 - \sqrt{\frac{2}{3}} \,\varphi + \frac{1}{3} \,\varphi^2 + {\cal O} (\varphi^{3}) \right) + \left( f_\kappa \,\tilde{\Lambda} - \frac{1}{\alpha} \right) \left( 1 - \sqrt{\frac{8}{3}} \,\varphi + \frac{4}{3} \,\varphi^2 + {\cal O} (\varphi^{3}) \right) \right] \nonumber \\
	&\simeq \,\left[ \widetilde {\cal D} + f_\kappa \,\tilde{\Lambda} + \sqrt{\frac{2}{3}} \left( \frac{1}{\alpha} - 2 f_\kappa \,\tilde{\Lambda} \right) \varphi + \:\frac{1}{3} \left( - \frac{3}{\alpha} + 4 f_\kappa \,\tilde{\Lambda} \right) \varphi^2 + {\cal O} (\varphi^{3}) \right]~.
\end{align}
We shall now proceed to demonstrate that a purely D-particle recoil driven slow-roll inflation is impossible, in the case of weak condensate fields. To this end, one needs to study the slow-roll conditions and the WMAP imposed constraints \cite{Komatsu:2010fb, Hinshaw:2012aka}. More precisely, the number $N$ of e-folds, the slow-roll parameters $\epsilon, \eta, \xi$, the spectral index $n_{\rm s}$ and the WMAP normalisation read
\begin{align} \label{srp}
	N &\equiv - \int_{\varphi_i}^{\varphi_e} \frac{V}{V'} \;\mathrm{d} \varphi \simeq 60 ~, \nn
	\epsilon &\equiv \frac{1}{2} \,M_{\rm{Pl}}^2 \left( \frac{V'}{V} \right)^2 \ll 1 ~, \qquad \eta \equiv M_{\rm{Pl}}^2 \left( \frac{V''}{V} \right) \ll 1 ~, \qquad \xi \equiv M_{\rm{Pl}}^4 \left( \frac{V''' \,V'}{V^2} \right) \ll 1 ~, \nn
	n_{\rm s} &\equiv 1 - 6 \epsilon + 2 \eta \simeq 0.96 ~, \qquad \left( \frac{\frac{1}{\kappa_{\rm{eff}}^2} \,V}{\epsilon} \right)^{\sfrac{1}{4}} \!= 0.0275 \;M_{\rm Pl} ~.
\end{align}
We note first that, as a result of the dilaton equation, the condition eq.~(\ref{negLambda}) is imposed, implying that $\tilde \Lambda$ is negative and the coefficients of the linear and quadratic terms in $\varphi$ that are present in eq.~(\ref{potential}) are non-zero. This leads to non-trivial slow-roll parameters, yielding
\begin{align}
	N &= \frac{V}{V'} \,\Delta \varphi \qquad \Rightarrow \qquad \epsilon = \frac{\left( \Delta \varphi \right)^2}{2 \,N^2} ~, \qquad \eta = \frac{V''}{V'} \frac{\Delta \varphi}{N}~, \nonumber
\end{align}
which, combined with the definition of $n_{\rm s}$, leads to
\begin{align}
	n_{\rm s} = 1 - 6 \,\frac{\left( \Delta \varphi \right)^2}{2 \,N^2} + 2 \,\frac{V''}{V'} \frac{\Delta \varphi}{N} \qquad \Leftrightarrow \qquad \frac{V''}{V'} = \frac{3}{2} \frac{\Delta \varphi}{N} - \frac{N}{\Delta \varphi} \,\frac{1 - n_s}{2}~.
\end{align}
Since $N \sim 60$ and $\Delta \varphi \ll 1$, one has $(1 - n_{\rm s}) \,N / \Delta \varphi \gg 1$ and $\Delta \varphi / N \ll 1$, hence to a good approximation
\begin{align}
	\frac{V''}{V'} \simeq - \frac{N}{\Delta \varphi} \,\frac{1 - n_{\rm s}}{2} \label{VsecVpEta} \quad &\Leftrightarrow \quad \frac{\frac{2}{3} \left( - \frac{3}{\alpha} + 4 f_\kappa \,\tilde{\Lambda} \right)}{\sqrt{\frac{2}{3}} \left( \frac{1}{\alpha} - 2 f_\kappa \,\tilde{\Lambda} \right)} \simeq - \frac{N}{\Delta \varphi} \,\frac{1 - n_{\rm s}}{2} \nn
	&\Leftrightarrow \qquad \frac{1}{\alpha} \left( -3 + \sqrt{\frac{3}{2}} \,\frac{N}{\Delta \varphi} \,\frac{1 - n_{\rm s}}{2} \right) \simeq - 2 f_\kappa \,\tilde{\Lambda} \left( 2 - \sqrt{\frac{3}{2}} \frac{N}{\Delta \varphi} \,\frac{1 - n_{\rm s}}{2} \right)~.
\end{align}
Since $\sqrt{\frac{3}{8}} \,N \left( 1 - n_{\rm s} \right) \sim 1.5$ and $\Delta \varphi \ll 1$, the constant term in each bracket in the above equation is small and can be neglected. This yields
\begin{align}
	 \frac{1}{\alpha} \left( \frac{1.5}{\Delta \varphi} \right) \simeq 2 f_\kappa \,\tilde{\Lambda} \left( \frac{1.5}{\Delta \varphi} \right) \quad \Leftrightarrow \quad \frac{1}{\alpha} \simeq 2 f_\kappa \,\tilde{\Lambda} \simeq 2 \,\kappa_{\rm eff}^2 \,\Lambda_0 ~,
\end{align}
leading to a positive $\Lambda_0$, incompatible with eq.~(\ref{negLambda}), which is required by the dilaton equation.

One is then led to the conclusion that no valid scenario exists for small field inflation induced by D-particles alone.

%----------------------------------------------------------------------------------%
\subsection{Inflation for large recoil velocity condensate fields}

Let us now concentrate on the low string scale case, where the condensate (\ref{cond}) is large
\be \label{lss}
	M_{\rm s} \ll H_I \sim 10^{-5} \,M_{\rm Pl} \ll M_{\rm Pl} ~,
\ee
where we used that the Planck data~\cite{Planck} point towards the fact that $H_I \simeq 10^{-5} \,M_{\rm Pl}$. In this case, one cannot expand the square root of the Born-Infeld action, but one can approximate it by ignoring the constant inside, that is the (\emph{Euclideanised} space-time) action (\ref{action1}) becomes (setting $\phi_0 = 0$ from now on)
\begin{equation} \label{action2}
	S_{\rm eff~4D} \simeq \int d^4x \,\sqrt{g} \left[ -\frac{1}{4} \,{\mathcal G}_{\mu\nu} {\mathcal G}^{\mu\nu} - \frac{T_3}{g_{{\rm s}0}} \,\sqrt{\frac{{\mathcal C}^{\mathcal E}}{2}} - \frac{{\tilde \Lambda}}{\kappa_0^2} + \frac{1}{\kappa_0^2} \left( 1 + \kappa_0^2 \,\frac{\alpha \,T_3}{g_{{\rm s}0}} \,\sqrt{\frac{{\mathcal C}^{\mathcal E}}{2}} \right) R(g) \right]
\end{equation}
where $\alpha$ is given by eq.~(\ref{defa}) and the condensate is positive. Let us define the dimensionless field
\begin{equation} \label{defsigma}
	\sigma (t) \equiv \kappa_0^2 \,\frac{\alpha \,T_3}{g_{{\rm s}0}} \,\sqrt{\frac{{\mathcal C}^{\mathcal E}(t)}{2}} ~> 0 ~,
\end{equation}
by means of which the action (\ref{action2}) becomes
\be \label{action3a}
	S_{\rm eff~4D} \simeq \int {\rm d}^4x \,\sqrt{g} \,\frac{1}{\kappa_0^2} \left[ - \frac{\kappa_0^2}{4} \,{\mathcal G}_{\mu\nu} \,{\mathcal G}^{\mu\nu} - {\tilde \Lambda} - \frac{\sigma}{\alpha} + \left( 1 + \sigma \right) R(g) \right] ~.
\ee
Before going further, we should make some remarks regarding the magnitude of the condensate field (\ref{defsigma}). First of all we observe that, if we were to follow the galactic scale lensing analysis in section~\ref{sec:lensing} using brane tensions that satisfy eq.~(\ref{setting})~\cite{msy}, the condensate field would be small $\sigma (t) \ll 1$ since
\be \label{small}
	\sigma (t) \simeq \frac{g_{{\rm s}0}}{2 \sqrt{3}} \left( \frac{H_I}{M_{\rm Pl}} \right)^2 \ll 1~,
\ee
where we have used eqs.~(\ref{cond}) and (\ref{egc}) and the fact that $M_{\rm s} \ll M_{\rm Pl}$ (cf. eq.~(\ref{lss})) to approximate $M_{\rm Pl}^2 \simeq \kappa_0^{-2} $. As we saw in subsection~\ref{sec:smallcondinfl}, such weak condensates cannot lead to slow-roll inflation.

Here we are interested in large field inflation, which, as we shall demonstrate below, can be induced by large recoil-velocity condensate fields $\sigma (t) \gg 1$. The latter condition may be achieved if we relax eq.~(\ref{setting}) and use large brane tensions, namely
\be \label{lbt}
	\frac{(2\pi \alp)^2 \,T_3}{g_{{\rm s}0}} \equiv \mathcal{J} \gg 1 ~.
\ee
The reader should recall eq.~(\ref{defJ}) where the parameter $\mathcal{J}$ was first defined. In such a case, we obtain from eq.~(\ref{defsigma})
\be \label{lc}
	\sigma(t) \simeq \frac{g_{{\rm s}0}}{2 \sqrt{3}} \,\mathcal{J} \,\kappa_0^2 \,H_I^2 \simeq \frac{g_{{\rm s}0}}{\sqrt{3}} \,\mathcal{J} \left( \frac{H_I}{M_{\rm Pl}} \right)^2~,
\ee
which can be much larger than one. In this case from eq.~(\ref{egc}) we obtain that $M_{\rm Pl} ^2 = \frac{1}{\kappa_0^2} + \frac{\mathcal{J}}{24} \,M_{\rm s}^2$ and $\kappa_0^2$ is a parameter independent from $M_{\rm Pl}$. Without loss of generality one can simply assume the relation (\ref{choice}), used in the previous section for the lensing analysis, which is consistent with eq.~(\ref{lss}). In such a case one obtains
\be \label{ivs}
	\sigma (t) \sim 8 \sqrt{3} \,g_{{\rm s}0} \left( \frac{H_{\rm I}}{M_{\rm s}} \right)^2 \gg 1 ~,
\ee
during inflation. In the remainder of this subsection we shall stick to this case.

We now proceed to discuss how inflation is induced by such large condensates. As we shall show, the induced inflation is of Starobinsky-type~\cite{dflation}. We first redefine the metric in eq.~(\ref{action3a}) as
\be \label{tildemetr}
	g_{\mu\nu} \rightarrow \tilde g_{\mu\nu} = (1 + \sigma ) g_{\mu\nu} ~.
\ee
We also define a canonically normalised scalar field
\be \label{cnsf}
	\varphi (t) = \sqrt{\frac{3}{2}} \,{\rm ln}\Big(1 + \sigma (t)\Big) ~,
\ee
in terms of which the action (\ref{action3a}) becomes
\begin{equation} \label{action3}
	S_{\rm eff~4D} \simeq \int d^4x \sqrt{\tilde g} \,\frac{1}{\kappa_0^2} \left[ R(\tilde g) + \frac{1}{2} \,\partial_\mu \varphi \,\partial^\mu \varphi - \frac{\kappa_0^2}{4} \,{\mathcal G}_{\mu\nu} {\mathcal G}^{\mu\nu} - \frac{e^{-\sqrt{\frac{2}{3}} \,\varphi}}{\alpha} - \left( \tilde \Lambda - \frac{1}{\alpha} \right) e^{-2 \sqrt{\frac{2}{3}} \,\varphi} \right]
\end{equation}
where we took into account the conformal nature of the flux gauge term in four space-time dimensions. We may assume now that the flux field condenses into a constant one, which contributes to the vacuum energy as
\be \label{fluxcond}
	\frac{1}{4} \llang {\mathcal G}_{\mu\nu} \,{\mathcal G}^{\mu\nu} \rrang \equiv {\mathcal D} ~.
\ee
The last three terms in the Euclidean effective action (\ref{action3}), define the \emph{Euclideanised} (superscript ${\mathcal E}$) effective potential of the $\varphi$ field in the region of large values (defined with dimensions of $[mass]^2$)
\begin{equation} \label{poteucl}
	V^{\mathcal E} = - \kappa_0^2 \,\mathcal D - \frac{e^{-\sqrt{\frac{2}{3}} \,\varphi}}{\alpha} - \left( \tilde \Lambda - \frac{1}{\alpha} \right) e^{-2 \sqrt{\frac{2}{3}} \,\varphi} ~.
\end{equation}
The reader should take notice of the relative sign of the potential compared to the kinetic term of the scalar field in (\ref{action3}), as appropriate for a euclidean effective action, which is just the effective Hamiltonian of the system. We should now \emph{analytically} continue (\ref{poteucl}) back to the Minkowski space-time. This implies that, apart from the time being rendered a Minkowskian signature, $x^4 \to i t$, the field $\varphi$ acquires an imaginary part
\be \label{imaginary}
	\sqrt{\frac{2}{3}} \,\varphi \rightarrow {\rm ln}(i | \sigma |) = {\rm ln}|\sigma| + i \frac{\pi}{2} = \sqrt{\frac{2}{3}} \,\tilde \varphi + i \frac{\pi}{2} ~,
\ee
where now the field $\tilde \varphi$ is real.

Thus, from eq.~(\ref{poteucl}), and assuming that the flux condensate $\mathcal D$ is of ``electric'' type so that under analytic continuation to Minkowski space-time one has $\mathcal D \to -\mathcal D$, the potential acquires an imaginary part and is approximated by
\begin{align} \label{effpot}
	V(\varphi) \simeq \kappa_0^2 \,{\mathcal D} + \left( \tilde \Lambda - \frac{1}{\alpha} \right) e^{-2\sqrt{\frac{2}{3}} \,\tilde\varphi} + i \,\frac{e^{-\sqrt{\frac{2}{3}} \,\tilde\varphi}}{\alpha} ~.
\end{align}
Notice that the imaginary part of the potential is the only one appearing in the analytically-continued effective action (\ref{action3}), given that the kinetic terms are real, since the imaginary part of the $\varphi$ field is constant in space-time.

The presence of an imaginary part indicates an instability of the de Sitter inflationary vacuum which is not an unwelcome fact. The field will roll down towards smaller values of $H^2$. Eventually the condensate (\ref{cond}) will become smaller than the Born-Infeld critical field and hence the imaginary part will disappear. In this regime, one may expand the square root of the Born-Infeld action, as done in ref.~\cite{msy}, to obtain the effective action relevant for the radiation and matter eras.

The imaginary part of the potential gives by definition the width, or equivalently, the inverse of the lifetime of the de Sitter vacuum, namely
\be \label{lifetime}
	\tau = \hbar \,\Gamma^{-1} \sim \kappa_0^{-1} \,\alpha \,e^{\sqrt{\frac{2}{3}} \,\tilde \varphi}
\ee
which is sufficiently long (as compared to the reduced Planck time $\kappa_0$) for any positive value of $\tilde \varphi \sim \sqrt{6} \;{\rm ln}(H_{\rm I}/M_{\rm s})$.

The real part of the effective potential (\ref{effpot})
\begin{equation} \label{realeffpot}
	{\rm Re} \,V(\tilde \varphi ) = \tilde {\mathcal D} + \left( \tilde \Lambda - \frac{1}{\alpha} \right) e^{-2 \sqrt{\frac{2}{3}} \,\tilde\varphi}~, \quad \tilde{\mathcal D} \equiv \kappa_0^2 \,{\mathcal D}~,
\end{equation}
is of Starobinsky type, provided one can tune the flux-field condensate to be $\tilde {\mathcal D} > 0$ and such that the minimum of the potential occurs for the field value $\tilde \varphi = 0$ and corresponds to zero potential. The quantity $\tilde \Lambda$ is negative, as a consequence of the dilaton equation of motion, and in fact can be tuned to the value given by eq.~(\ref{t3ltilde}) in order to ensure continuity of the inflation phase with the growth era. Hence, the coefficient of the $e^{-2\sqrt{\frac{2}{3}} \,\tilde \varphi}$ term is negative relative to $\tilde{\mathcal D}$. An \emph{important feature} of the approach here is that it is the gauge field flux condensate ${\mathcal G}_{\mu\nu} {\mathcal G}^{\mu\nu}$ that induces a de Sitter phase (positive, almost constant, vacuum energy), and hence inflation, but it is the recoiling D-particles velocity vector field that induces a slowly rolling scalar degree of freedom that allows exit from inflation.

\begin{figure}[t] \begin{adjustwidth}{-5em}{-4em}
	\begin{center}
	\includegraphics[height=9.5cm, keepaspectratio]{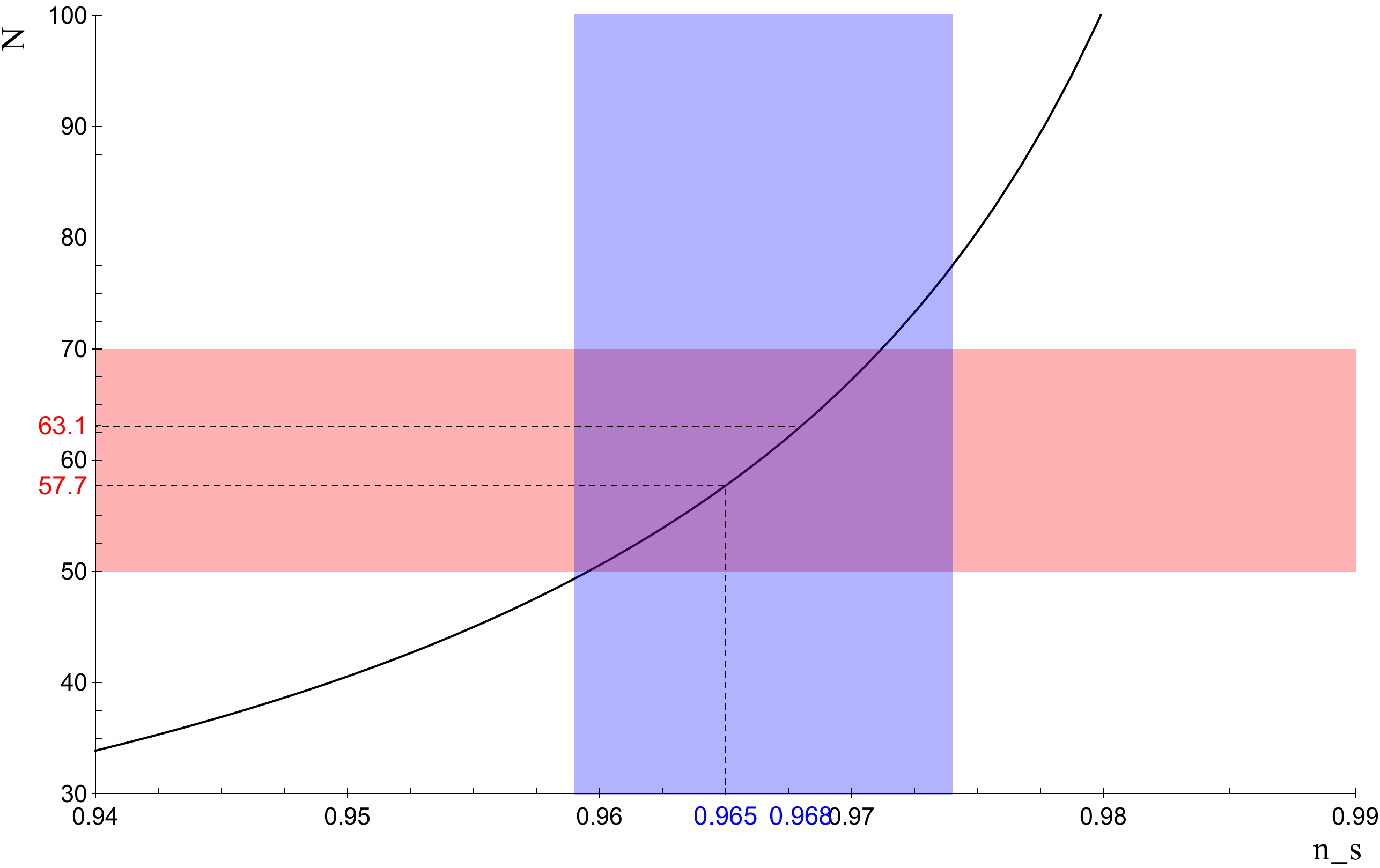}
	\caption{$N$ as a function of $n_{\rm s}$.}
	\label{ns(N)}
	\end{center}
\vspace{-1em}\end{adjustwidth} \end{figure}

Now, let us see how one can get slow-roll inflation in this case. We should recall that slow-roll requires the conditions (\ref{srp}), which we will now evaluate. For large condensate $\sigma$, one has
\begin{align}
	V(\tilde \varphi ) &= \tilde {\mathcal D} + \left( \tilde \Lambda - \frac{1}{\alpha} \right) e^{-2 \sqrt{\frac{2}{3}} \,\tilde \varphi} = \tilde {\mathcal D} - A\,e^{-B \tilde \varphi} ~, \nn
	V'(\tilde \varphi ) &= AB\,e^{-B \tilde \varphi} ~, \hspace{3em} V''(\tilde \varphi ) = - AB^2\,e^{-B \tilde \varphi} ~, \hspace{3em} V'''(\tilde \varphi ) = AB^3\,e^{-B \tilde \varphi} ~,
\end{align}
where we defined the constants $A \equiv - \left( |\tilde \Lambda| + \frac{1}{\alpha} \right) > 0$ and $B \equiv 2 \sqrt{\frac{2}{3}} \simeq 1.15$. Hence
\begin{subequations} \begin{align}
	N &\simeq \frac{\tilde {\mathcal D}}{AB^2} \,e^{B \tilde \varphi} \label{Nliteral} \\
	\epsilon &\simeq \frac{A^2 B^2}{2 \tilde {\mathcal D}^2} \,e^{-2 B \tilde \varphi} = \frac{1}{2B^2 N^2} \qquad \mathrm{and} \qquad \eta \simeq - \frac{AB^2}{\tilde {\mathcal D}} \,e^{-B \tilde \varphi} = - \frac{1}{N} \\
	&\qquad \Rightarrow \quad 1 - n_{\rm s} = 6 \epsilon - 2 \eta = \frac{3}{B^2 N^2} + \frac{2}{N} \label{trinomialN} \\
	\xi &\simeq \frac{A^2 B^4}{\tilde {\mathcal D}^2} \,e^{- 2 B \tilde \varphi} = \frac{1}{N^2} \\
	\left( \frac{\sfrac{1}{\kappa_0^2} \,V}{\epsilon} \right)^{\frac{1}{4}} &\simeq \left( \frac{\tilde{\mathcal D}}{\epsilon} \right)^{\frac{1}{4}} = 0.0275 \;M_{\rm Pl} \\
	&\qquad \Rightarrow \quad \tilde{\mathcal D} \simeq \left(0.0275 \right)^4 \,\epsilon \;M_{\rm Pl}^4 \simeq \frac{5.7 \cdot 10^{-7}}{2B^2 N^2} \;M_{\rm Pl}^4
\end{align} \end{subequations}
where the last equation comes from the WMAP constraint. So, as is standard in Starobinsky-type inflation, the constant $A$ is not constrained by the slow-roll conditions (in our microscopic model, as we have already mentioned, we may tune it to the value determined by eq.~(\ref{t3ltilde}) by demanding continuity of the inflationary epoch to the galactic-growth era of the string universe). Thus, fixing $n_{\rm s}$ fixes $N$ (and vice-versa). Indeed, one gets (from solving the second degree trinomial in $N$ from eq.~(\ref{trinomialN}) and choosing the positive solution)
\be
	N = \frac{1}{1-n_{\rm s}} \left( 1 + \sqrt{1 + \frac{3 \left( 1 - n_{\rm s} \right)}{B^2}} \right) ~.
\ee
Planck 2015 analysis~\cite{Planck_infl} gives $n_{\rm s} = 0.968 \pm 0.006$ ($68 \%$ CL, PlanckTT+LowP) which is shifted towards higher values compared to earlier results, that gave a central value $n_{\rm s} =0.965$. The solid black line in figure~\ref{ns(N)} shows $N$ as a function of $n_{\rm s}$ for $B = 2 \sqrt{\sfrac{2}{3}}$. The vertical blue shaded area corresponds to the $68 \%$ CL interval for $n_{\rm s}$ corresponding to the two central values $n_{\rm s} = 0.965$ and $n_{\rm s} = 0.968$, while the horizontal shaded area in red shows the relevant interval for $N$. The black dashed line highlights the central values for $n_{\rm s}$ and the corresponding values for $N$. One notes the excellent fitting of the predictions of the model to the data.

If we adopt $n_{\rm s} = 0.965$, we get $N = 57.7$, leading to
\begin{subequations} \begin{align}
	\epsilon &\simeq 5.6 \cdot 10^{-5} \ll 1 ~, \qquad \eta \simeq - 1.7 \cdot 10^{-2} \ll 1 ~, \qquad \xi \simeq 3.0 \cdot 10^{-4} \ll 1 \\
	{\mathcal D} &\simeq 3.2 \cdot 10^{-11} \;M_{\rm Pl}^4 \qquad \Leftrightarrow \qquad \tilde {\mathcal D} \simeq 3.2 \cdot 10^{-11} \;M_{\rm Pl}^2~,
\end{align} \end{subequations}
showing that every constraint related to the slow-roll conditions is satisfied: $N \sim 60$, $\epsilon \ll 1$, $\eta \ll 1$, $\xi \ll 1$, while the WMAP constraint on $\sfrac{V}{\epsilon}$ gives constraints on the size of the potential.

The constant $A$ and the value of the field $\tilde{\varphi}$ are not constrained at all so far except in that their combination $A\,e^{-B \tilde \varphi}$ must satisfy
\be
	\tilde {\mathcal D} \gg A\,e^{-B \tilde \varphi} ~,
\ee
in order to ensure slow-roll. From eq.~(\ref{Nliteral}), we know
\be
	A \,e^{-B \tilde \varphi} = \frac{\tilde {\mathcal D}}{B^2 N} \simeq \frac{\tilde {\mathcal D}}{1.5 \cdot 10^2} \ll \tilde {\mathcal D} ~,
\ee
which confirms the consistency of our model.

Note that this result is not very sensitive to the value of $B$, especially to larger $B$. Indeed, even with $B$ a hundred times larger, $N$ is shifted towards lower values by less than $1$ unit. Lowering $B$ would modify a bit more our results but not much either, since with a $B$ twice as small, $N$ increases by less than $2$ units while with a $B$ four times smaller, $N$ increases by less than $8$ units (reaching about $70$ for $n_{\rm s} = 0.968$).

At this point it is worth making \emph{a few important remarks}. The considerations leading to eq.~(\ref{realeffpot}) indicate the possibility of inflationary scenarios for the redefined metric $\tilde g_{\mu\nu}$ through eq.~(\ref{tildemetr}). Nevertheless, for a slowly rolling condensate $\sigma(t)$, as required during inflation, the original metric is also inflationary, up to a time coordinate change. Indeed, let us denote by $\tilde t $ the cosmic time coordinate in the $\tilde g$-metric. The latter during inflation corresponds to a line-element of the form
\be
	d\tilde s^2 = d{\bar t}^2 - \tilde a (\bar t)^2 h_{ij}(x^k) \,dx^i dx^j = (1 + \sigma (t)) \,dt^2 - (1 + \sigma(t)) \,a(t)^2 h_{ij}(x^k) \,dx^i dx^j ~,
\ee
in a standard FLRW notation, where
\be \label{coordinates}
	d\tilde t = \sqrt{(1 + \sigma(t))} \,dt \simeq \sqrt{\sigma} \,dt ~, \qquad \tilde a (\tilde t) = \sqrt{(1 + \sigma (t))} \,a (t) \simeq \sqrt{\sigma} \,a(t)~, \qquad \sigma \gg 1 ~.
\ee
For a slowly moving $\sigma$ field (almost constant), the two metrics differ by an overal scale factor, and the corresponding Hubble parameters are related as follows (quantities with a tilde pertain to the metric (\ref{tildemetr}) and the coordinates (\ref{coordinates}))
\be
	{\tilde H} \equiv \frac{1}{\tilde a(\tilde t) } \,\frac{d}{d\tilde t} \,\tilde a (\tilde t) = \frac{1}{\sqrt{\sigma}} \left( \frac{\dot \sigma}{\sigma} + \frac{\dot a}{a} \right) \simeq \frac{1}{\sqrt{\sigma}} \,H~,
\ee
since during inflation $\dot \sigma /\sigma \ll 1$ and can be neglected in front of the $H=\dot a / a$ term (the overdot denotes time derivatives with respect to the cosmic time pertaining to the initial metric $g_{\mu\nu}$).

Taking into account (\ref{ivs}) as a concrete example, for $M_{\rm s} \ll H_I$, we can estimate
\be
	\tilde H \equiv H_I \sim \frac{1}{\sqrt{8 \sqrt{3} \,g_{{\rm s}0}}} \,\frac{M_{\rm s }}{H_{\rm I}} \,H~,
\ee
from which it follows that the inflationary scale of the original metric is much higher than $H_{\rm I}$, of order
\be \label{physical}
	H \sim \sqrt{8 \sqrt{3} \,g_{{\rm s}0}} \,\frac{H_{\rm I}^2}{M_{\rm s}} \gg H_I ~, \qquad M_{\rm s} \ll H_{\rm I}~. 
\ee
The reader should bear in mind that above we matched the cosmological observations~\cite{Planck} on inflation with predictions made by the conformal rescaled metric $\tilde g_{\mu\nu}$ (\ref{tildemetr}), and therefore it is the scale $\tilde H$ that we call the ``physical'' Hubble scale $H_{\rm I} \sim 10^{-5} \,M_{\rm Pl}$ used in observations~\cite{Planck}. For a smooth connection with the galaxy data in this case we should use the action (\ref{action3}) with the metric (\ref{tildemetr}), after the inflaton decays, that is we should couple it to matter and radiation. At the end of inflation the condensate of D-particles vanishes $\sigma \to 0$ and the two metrics coincide. At the radiation era that succeeds the exit from inflation, the condensate field $\sigma$ is replaced by the weak field $\widetilde F_{\mu\nu} \widetilde F^{\mu\nu}$ and the action (\ref{action2b}) describes now the dynamics. The only difference from the discussion in section~\ref{sec:lensing} occurs in the large brane tension eq.~(\ref{lbt}) which replaces eq.~(\ref{setting}). This will have quantitative only consequences in the allowed range of the $\beta$ parameter, and hence the density of D-particles relevant for growth and dark matter. From the lensing equation (\ref{lens2}), for instance, we see that it is now the parameter $\mathcal{J} \,|\beta|$ that is required to match the dark matter in a galaxy, and it is $\mathcal{J} \,|\beta|$ that appears on the left hand side of the inequality (\ref{upper}). This leads to much smaller upper bounds for the density of D-particles in order for them to mimic dark matter through their recoil-velocity-field fluctuations. In a similar manner, by repeating the analysis of ref.~\cite{msy} in this case, one arrives at much smaller lower bounds for the density of D-particles required for growth of structure in the D-material universe.
 
Another important aspect is that the condensate $\sigma(t)$ defined in eq.~(\ref{defsigma}), upon the redefinition of the metric (\ref{tildemetr}), can be expressed in terms of $\tilde g_{\mu\nu}$ as follows
\begin{align} \label{condtilde}
	\sigma = (1 + \sigma)^2 \tilde \sigma
\end{align}
where $\tilde \sigma $ is the condensate $\llang F_{\mu\nu} F_{\alpha\beta} \,{\tilde g}^{\mu\alpha} \,{\tilde g}^{\nu\beta} \rrang$, which, because is a scalar, will assume the same value if one passes onto the coordinates (\ref{coordinates}). For large $\sigma \gg 1$ one obtains from (\ref{condtilde}) $\sigma \sim 1/\tilde \sigma \gg 1$. In terms of the $\tilde \sigma$ field, the canonically normalized inflaton field defined in eq.~(\ref{cnsf}) reads $\varphi \sim -{\rm ln}(\tilde \sigma)$.

Finally, before closing this subsection, it is important to comment on the order of magnitude of the statistical parameter $\sigma_0^2 \equiv \sigma_{0,{\rm infl}}^2$ (\ref{fcond}) during the large-condensate inflationary era with $M_{\rm s} \ll M_{\rm Pl} = \kappa_{\rm eff}^{-1}$. As we have already mentioned, this parameter is assumed constant during inflation, due to the fact that during that epoch the brane world moves in a bulk region which is densely populated by D-particle defects, in such a way that there is a large incoming flux of D-particles from the bulk onto the brane compensating any potential dilution of their population on the brane due to the brane-universe expansion. From eqs.~(\ref{variance}), (\ref{ivs}) we estimate
\be
	\frac{\sigma_{0, \rm infl}^2}{\sigma_{\rm slow-roll}} \simeq \frac{g_{{\rm s}0}}{8 \pi^2 \sqrt{3}}~,
\ee
where $\sigma_{\rm slow-roll} = 8 \sqrt{3} \,g_{{\rm s}0} \left( \sfrac{H_{\rm I}}{M_{\rm s}} \right)^2$. For $M_{\rm s} = \mathcal{O}(10^4)$~GeV, we obtain $\sigma_{0,{\rm infl}}^2 = {\mathcal O}(10^{16})$.\footnote{Notice that for large condensates the restriction (\ref{smallf}) does not apply.}

%----------------------------------------------------------------------------------%
\subsection{Estimates of the age of the D-material universe}

Before closing this section we would like to make some crude estimates of the age of the D-material universe $t_0$ in the case where inflation is driven by strong condensates of D-particle recoil velocities. It goes without saying that, without detailed microscopic models it is not possible to find a precise connection of the value of $\sigma_0^2 \simeq |\beta|$ during galactic era with $\sigma_{0,{\rm infl}}^2$, which would allow for a precise estimate of $t_0$. Indeed, this would require knowledge of the bulk distribution of D-particles from the moment of the exit from inflation until the current era. In our complicated dynamical system, the equation of state of the pertinent cosmic fluid is not a constant and depends on many factors, including the density profile of the bulk and brane D-particles at any given era. Nevertheless, as we shall discuss below, one can make some simplifying assumptions, which allow us to make some estimates of the age of the D-material universe in a phenomenological context.

To this end, we first recall~\cite{msy} that in the case of complete dominance of the classical recoil-velocity condensates, based on statistical populations of D-particles whose dynamics is governed by a Born-Infeld action of the vector field alone, the equation of state of the recoil-velocity fluid would be $w=-1/3$. This is the limiting case (from above) for which acceleration of the universe occurs. From the corresponding Friedmann equation, that would lead to a linearly expanding universe with the cosmic time, $a(t) \sim t$, which is not physical. However, the D-material universe's Lagrangian is much more complicated than a simple Born-Infeld fluid. The presence of matter as well as non minimal couplings of the Born-Infeld factors with space-time curvature (cf. eq.~(\ref{action1})) alter the situation drastically and one expects, as already mentioned, a time (redshift) dependent equation of state for the total fluid, D-particles and matter strings, $w_{\rm total}(z)$, whose form is currently difficult to estimate without detailed knowledge of the density profile of the bulk D-particles.

Matter domination era in our case includes contributions to the stress tensor coming from the recoil velocities of the D-particles bound on the brane world, as a consequence of their interactions with string matter. Matter dominance, therefore, does not exclude the possibility that the contributions of the recoil-velocity fluid to the total energy density of the universe are of the same order of magnitude as the corresponding matter energy density during the galaxy formation era, that is with redshifts $z \gsim 1$, which corresponds to the upper bound in the inequality (\ref{upper}). On the other hand, for the current era, i.e. redshifts $z \lsim 10^{-2}$, current data indicate a cosmological constant dominance, $w_{\rm total}(z \simeq 0)=-1$. If we make the (considerable) simplification that there is a depletion of bulk D-particles from the exit of inflation era until the galaxy formation era, and if we assume a \emph{simple} power scaling of the energy density of the total fluid $\rho_{\rm total}$ with the scale factor of the universe as
\be \label{simple}
	\rho_{\rm total} \sim a^{-3(1 + w_{\rm total})}
\ee
with $w_{\rm total}$ approximately constant, then, from the Friedmann equation, we would obtain the following time dependence of the scale factor on the cosmic time
\be \label{totalfluid}
	a(t)_{\rm matter+D-particle-recoil-fluid} \sim t^{\frac{2}{3(1 + w_{\rm total})}}~.
\ee
Then, assuming that eq.~(\ref{stat}) is valid from the galactic era all the way back to the exit from inflation epoch, we obtain that, for a given moment $t$ in the history of the D-material universe, the statistical variance of the D-particle recoil velocities behaves as
\be \label{statt}
	\llang u_i u^i \rrang \equiv \sigma_0^2 (t) = \frac{|\beta|}{(\sfrac{a(t)}{a_0})^3}
\ee
where $a_0 \equiv a(t_0)$, with $t_0$ the universe age, is today's value of the scale factor, which is re-instated here for reasons that will become clear below.

At the exit from inflation, at cosmic times $t=t_{\rm infl}$, the variance is assumed to have the value $\sigma_{0,{\rm infl}}^2$. Since we are interested in an order of magnitude estimate of the universe age, we are at liberty to ignore the short duration of the late de Sitter accelerating phase of the universe, and assume the scaling (\ref{simple}) for the statistical fluctuations of the recoil velocities from the inflationary era till practically today $t=t_0$. Then from eq.~(\ref{statt}) we obtain
\be \label{sigmas}
	\sigma_{0,{\rm infl}}^2 \,a(t_{\rm infl})^3 \sim |\beta| \,a_0^3 ~.
\ee
Using the upper and lower bounds for $|\beta|$ given in eqs.~(\ref{upper}) and (\ref{minbeta}), so that the recoil velocity fluid either mimics the dark matter in galaxies (upper bound) or at least is responsible for inducing growth of structures in the universe (lower bound), we obtain the following allowed range for the D-material universe age
\begin{align} \label{duage}
	\left( 10^{60} \,\frac{g_{{\rm s}0}}{\pi} \,\frac{H_{\rm I}}{M_{\rm Pl}} \,\mathcal H \right)^{1 + w_{\rm total}} &\lesssim \frac{t_0}{t_{\rm infl}} \lesssim \left( 10^{61.5} \,\frac{g_{{\rm s}0}}{\pi} \,\frac{H_{\rm I}}{M_{\rm Pl}} \right)^{1 + w_{\rm total}} \nn
	\Rightarrow \quad 10^{53.5 \,(1 + w_{\rm total})} &\lesssim \frac{t_0}{t_{\rm infl}} \lesssim 10^{55 \,(1 + w_{\rm total})}~,
\end{align}
where in the last line we used that $H_{\rm I} \simeq 10^{-5} M_{\rm Pl}$, $g_{{\rm s}0} \sim 0.1$, $\pi \sim 10^{0.5}$ and $\mathcal H = {\mathcal O}(1)$. Using that in conventional cosmology one estimates that exit from inflation occurs at times $t_{\rm infl} \sim 10^{12} \, t_{\rm Pl}$, where $t_{\rm Pl}$ is the Planck time ($10^{-45}$~s), we observe that the age of the universe in units of Planck time is estimated to be
\be \label{age}
	10^{65.5 \,(1 + w_{\rm total})} \lesssim \frac{t_0}{t_{\rm Pl}} \lesssim 10^{67 \,(1 + w_{\rm total})} ~.
\ee
If we insist that the age of the D-material universe $t_0$ is in agreement with the corresponding $\Lambda$CDM model estimates from Planck data~\cite{Planck}, i.e. $t \sim 10^{60} t_{\rm Pl}$, then in case the upper bound (\ref{upper}) is satisfied --- that is when the D-particle recoil-velocity fluid mimics dark matter in the galaxies, as far as lensing is concerned --- we obtain for the equation of state $w_{\rm total} = -0.085$. On the other hand, for the lower bound case (\ref{minbeta}) to be satisfied --- that is when the fluid of recoiling D-particles induces growth of structures in the universe, but falls short of reproducing the lensing effects of dark matter in galaxies --- one obtains $w_{\rm total} =-0.105$.

These values are not far from the pure Born-Infeld equation of state $w=-1/3$ for classical condensates~\cite{msy}. However, given that $w=-1/3$ is the maximum value for inducing acceleration in an Einstein universe, that satisfies the positive energy conditions, i.e. $w > -1$, we observe that in the matter dominated era the D-material universe decelerates, as it should be. As discussed in refs.~\cite{odintsov} and \cite{msy}, quantum fluctuations of the recoil-velocity condensates that satisfy Born-Infeld dynamics, when they are dominant, can lead to an accelerating almost de Sitter phase, with equation of state near to $p \simeq -\rho$. To match with the current universe phenomenology, according to which the universe today appears to be in a de-Sitter-like phase, one is led to the conclusion that quantum fluctuations of the weak condensates, that characterise the current era, dominate over the classical statistical effects. This is plausible in the low temperatures of the current universe.

%%%%%%%%%%%%%%%%%%%%%%%%%%%%%%%%%%%%%%%%%%%%%%%%%%%%%%%%%%
\section{Conclusions and outlook} \label{sec:conclusions}

In this work we built upon our previous discussion on the potential r\^ole of the recoil-velocity fluctuations of D-particle (effectively point-like) defects in brane universes, by presenting a cosmic evolution of the so-called D-material universe. The latter is a brane world which is punctured by populations of D-particles and propagates in a bulk space with varying densities of such defects. In the early stages of the universe, one may encounter dense populations of bulk D-particles, which imply a dense population of D-particles bound to the brane world. For low string scales $M_{\rm s}$ compared to the Hubble scale and for sufficiently large brane tensions compared to $M_{\rm s}^4$, the recoil velocity fluctuations (as a result of the interactions of D-particles with open strings representing matter and radiation on the brane world) lead to the formation of large condensate scalar fields that can drive inflation. In the case of large string scales compared to the Hubble scale, or smaller brane tensions of order of $M_{\rm s}^4$, the resulting condensates are small and cannot drive inflation. In such, inflation might be induced by other mechanisms, for instance it may be driven by large negative values of a slowly rolling dilaton field.

As (the cosmic) time lapses, the universe exits from a bulk region of such dense D-particle populations, inflation ends and the universe enters a radiation dominated era, with power law expansion of the scale factor in cosmic time. In such a case, the recoil velocity fluctuations of the D-particle diminish with the inverse cubic power of the scale factor. The pertinent condensates are weak. At such late eras of the universe, it has been shown that the recoil-velocity-fluctuation fluid may ``mimic'' dark matter in a way compatible with lensing phenomenology. Of course, given that the underlying string theory contains its own particle dark matter candidates, our findings here should be interpreted only as suggesting that the recoil velocity component might be the dominant one in agreement with current lensing data. Certainly, given that this assumption implies only upper bounds for the pertinent densities of D-particles, the picture of a multicomponent dark matter where both conventional particle and D-particle candidates might play an equal r\^ole in dark matter composition cannot be excluded at present. The density of D-particles at a given era in the history of the D-material universe is in a sense a free parameter in our low-energy treatment, although this can be actually controlled by performing numerical simulations of the evolution of a dense population of D-particles in colliding brane scenarios (whereby the collision implies the initial big-bang-like cosmically catastrophic event~\cite{dfoam}). This is not feasible at present.

Before closing we should point out that there are some important predictions of the D-material universe, given that the presence of populations of defects breaks translational invariance as well as Lorentz symmetry, the latter though only through quadratic recoil velocity fluctuations (on average the linear recoil velocity effects have been assumed to vanish (\ref{stat})). The populations of D-particle defects act as a kind of ``medium'' (\emph{D-foam}~\cite{dfoam}) which has a \emph{non-trivial refractive} index, that leads to time delays of the more energetic particles that interact non trivially with the D-particle defects. String theory considerations~\cite{refractive} imply different (and much more pronounced) refractive index for particles neutral under the standard model group as compared with the rest of the probes. In this sense, the dominant effects of the D-particle medium are expected to characterise photons and gravitons. In the early universe, where the density of D-particles are significantly higher than that in the current era, one expects that such Lorentz-violating effects of the D-foam affect significantly the propagation of primordial gravitational waves. In our inflationary phenomenology above we assumed standard analysis of the cosmological perturbations in order to match the slow roll parameters to the data, and in particular the tensor to scalar ratio. In the actual situation, where the dynamics of the densely populated medium of D-particles is properly taken into account, one may have some non trivial effects on this ratio, which might lead to observational signatures. This is an open issue that we plan to pursue in the future.

A final comment concerns the production of D-particles, in case their masses are less than 7 TeV, in the run II of the Large Hadron Collider (LHC) and their potential detection. As discussed in ref.~\cite{mitsou}, production of neutral D-$\overline{\rm D}$ pairs, from decays of highly-energetic off-shell $Z^0$-bosons, is a rare but possible event at LHC. Such production schemes follow from generic properties of \emph{D-matter} that characterise its coupling with ordinary Standard Model particles~\cite{dmatter}. The neutral defect pairs should manifest themselves in a way similar to ordinary particle/antiparticle dark-matter pairs at colliders. However, the D-particles have an additional peculiar property, which implies non conventional ways of detection. Specifically, their presence results in a deficit angle in the neighboring space-time~\cite{acs}, in a remote analogy to the effects (on the exterior space-time) of a global monopole, a configuration appearing in field-theoretic models with spontaneous breaking of rotational O(3) symmetry~\cite{globalmono}. Once therefore a D-particle is produced (in a pair with its antiparticle) in a collider, the colliding Standard-Model particles in the beam will find themselves in the environment of a space-time with a deficit angle. It is known that in such space-times scattering amplitudes produce local maxima (formally divergencies) when the scattering angle equals the deficit angle of the space-time~\cite{papavass}. This leads to strange scattering patterns (``\emph{Newton-like} rings'') around the trajectory of the defect, thereby making its detection possible in ATLAS and CMS LHC experiments, or even in the MoEDAL LHC experiment~\cite{moedal}. The latter is the seventh LHC recognised experiment, which is dedicated to the detection of highly ionising avatars of new physics, including the aforementioned D-matter, which though would not behave as highly ionising but result in these strange patterns of scattering of Standard Model matter in the environment of the defect. The latter are detectable in principle~\cite{dinmoedal} by the ``scars'' they may leave in the various types of passive detectors (such as TimePix) surrounding the collision point of the LHCb-experiment, near which MoEDAL is located. In view of the interesting cosmological properties of D-matter, outlined in the present work and in ref.~\cite{msy}, producing it at LHC, if it exists and is sufficiently light, would further enhance the opportunities of studying its peculiar properties and unravel its braney structure. This in turn, may result in a better understanding of fundamental properties of brane theory itself.

%%%%%%%%%%%%%%%%%%%%%%%%%%%%%%%%%%%%%%%%%%%%%%%%%%%%%%%%%%
\section*{Acknowledgements}

The work of T.E. is supported by a King's College London GTA Graduate studentship, while that of N.E.M. is supported in part by the London Centre for Terauniverse Studies (LCTS), using funding from the European Research Council via the Advanced Investigator Grant 267352, and by STFC (UK) under the research grant ST/L000326/1. We acknowledge discussions within the MoEDAL Collaboration, to which N.E.M. and M.S. belong, as well as within the Work Package 3 of the Theory Working Group of the Euclid Consortium, to which M.S. is a member.

%%%%%%%%%%%%%%%%%%%%%%%%%%%%%%%%%%%%%%%%%%%%%%%%%%%%%%%%%%
\appendix 

\section{Background field considerations \label{sec:bfe}}

Here, we discuss background field considerations which satisfy equations of motion obtained from the actions (\ref{action1}) and (\ref{actionSmallCondensate}). For the cosmological time scales we have been interested in from ref.~\cite{msy}, we have considered the FLRW space-time metric backgrounds
\begin{align} \label{flrwmetric}
	g^{\rm FRW}_{\alpha \beta} {\rm d}x^\alpha {\rm d}x^\beta = - {\rm d}t^2 + a^2(t) \left( {\rm d} r^2 + r^2 {\rm d}\theta^2 + r^2 \sin^2 \theta \,{\rm d}\varphi^2 \right) ~,
\end{align}
which, in conformal time $\eta$ used in our analysis below, yields
\begin{align} \label{ctm}
	g^{\rm FRW}_{\alpha\beta} {\rm d}x^\alpha {\rm d}x^\beta = - a^2(\eta) \,{\rm d}\eta^2 + a(\eta)^2 \left( {\rm d}r^2 + r^2 {\rm d}\theta^2 + r^2 \sin^2 \theta \,{\rm d}\varphi^2 \right) ~.
\end{align}
The dimensionful (dimension [mass]) cosmological form of the recoil vector field $A_\mu$ and its field strength $F_{\mu\nu} \equiv \partial_\mu A_\nu - \partial_\nu A_\mu$ on the D3 brane universe take the form
\begin{align} \label{vectorfield}
	A_i \equiv - \frac{1}{\sqrt{\alp}} \,a^2(t) \,u_i ~ , \qquad F_{0i} = - \frac{2}{\sqrt{\alp}} \,{\dot a} a \,u_i ~,
\end{align}
where the overdot denotes derivative with respect to the FLRW cosmic time $t$ and $\alpha^\prime$ is the Regge slope of the string (of dimension [length]$^2$).

This form of the vector field has been found by observing that, from the point of view of a $\sigma$-model perturbation, the latter corresponds to an ``impulse'' vertex operator on the world-sheet boundary $\partial \Sigma$ of the form~\cite{gravanis}
\begin{equation} \label{recvertex}
	V_\sigma^A = \frac{1}{2 \pi \,\alp} \,\oint_{\partial \Sigma} \,g_{ij} \,Y^j (t) \,\Theta (t - t_c) \,\partial_n X^i ~,
\end{equation}
where the integral is over the world-sheet boundary, $g_{ij}$ denotes the spatial components of the metric, $\partial_n $ is a normal world-sheet derivative, $X^i$ are $\sigma$-model fields obeying Dirichlet boundary conditions on the world sheet, and $t$ is a $\sigma$-model field obeying Neumann boundary conditions on the world-sheet, whose zero mode is the cosmic target time.

The path $Y^i(t)$ may be identified with the geodesic equation of a massive D-particle in the space-time described by the metric $g_{ij}$. The quantity $\Theta (t - t_{\rm c})$ is the Heaviside step function, expressing the instantaneous action (impulse) on the D-particle at $t=t_{\rm c}$. From a world-sheet viewpoint, the Heaviside function is an operator, which is such that the impulse/recoil operator (\ref{recvertex}) satisfies a logarithmic conformal algebra on the world-sheet of the string~\cite{gravanis}, which is the limiting case between conformal theories and general renormalisable two-dimensional theories, that can be classified by conformal blocks. For the purposes of this work and that of ref.~\cite{msy}, we shall work in times $t > t_{\rm c}$ so that the Heaviside function can be set safely to one.

Upon a T-duality transformation, assuming it to be an exact symmetry of the underlying string theory, we observe that the vertex operator (\ref{recvertex}) corresponds to that of a covariant vector (gauge) field\footnote{That there is an Abelian gauge symmetry associated with the vertex (\ref{tdual}) is obvious due to the fact that, upon a $U(1)$ target-space gauge transformation, with parameter $\theta(X^\alpha(\sigma,\tau)$), under which $A_\mu \to A_\mu + \partial_\mu \theta (X)$, the vertex remains unchanged, since $\oint_{\partial \Sigma} \,\partial_\mu \theta (X) \,\partial_\tau X^\mu = \oint_{\partial \Sigma} \frac{{\rm d}}{{\rm d}\tau} \theta = 0$, since the boundary of a boundary is zero by construction.}
\be \label{tdual}
	V_A = \frac{1}{2 \pi \,\sqrt{\alp}} \,\oint_{\partial \Sigma} \,{\rm d}\tau \,A_\mu \partial_\tau X^\mu~,
\ee
where $\partial_\tau$ denotes tangential world-sheet derivative. The vector field has spatial components~\cite{gravanis,msy}
\begin{equation} \label{vector}
	A_i = \frac{1}{\sqrt{\alp}} \,g_{ij} \,Y^j(t) \,\Theta (t-t_{\rm c}) ~.
\end{equation}
For a FLRW background with a power-law scale factor $a(t) \sim t^p$ relevant for the matter dominated era we were interested in ref.~\cite{msy}, and for times large compared to the moment of impact $t_{\rm c}$ for any given D-particle, one has~\cite{gravanis}
\begin{equation} \label{geodesics}
	Y^i(t) \simeq \frac{v^i}{1-2p} \left( t \,\frac{a^2(t_{\rm c})}{a^2(t)} - t_{\rm c} \right) + \dots \simeq - \frac{v^i}{1-2p} \,t_{\rm c} \qquad t \gg t_0~,
\end{equation}
which, on account of eq.~(\ref{vector}), implies the form (\ref{vectorfield}), having absorbed irrelevant numerical factors into the definition of the recoil velocity and restored the correct dimensionality via appropriate powers of $\sqrt{\alp}$.

The temporal component of the covariant vector field $A_0$ cannot be fixed in this approach, given that the target time coordinate satisfies Neumann boundary conditions on the world-sheet, and as such $\partial_n t (\sigma) = 0$ and thus does not appear in the original vertex (\ref{recvertex}) in the Dirichlet picture. In ref.~\cite{msy}, and for the case of cosmological space-times \emph{only}, we have covariantised the vector background by considering the temporal component of the (T-dual) vector field to be such that the four-vector field $A_\mu$ (of mass dimension one) assumes the form
\be \label{covariant}
	A_\mu = - \frac{1}{\sqrt{\alp}} \,g_{\mu\nu}(t) \,u^\nu ~,
\ee
where $u^\mu = \mathrm{d}x^\mu / \mathrm{d}\tau$ is a (dimensionless) four-velocity, satisfying the time-like constraint (in our convention)
\be \label{velconstraint}
	u^\mu \,u^\nu \,g_{\mu\nu} = - 1 ~.
\ee
The cosmological FLRW space-time have the conformally flat form (in conformal time $\eta$ coordinates, cf. (\ref{ctm}))
\be \label{conftimemetric}
	g_{\mu\nu} = C(\eta) \,\eta_{\mu\nu}~,
\ee
where $C(\eta)$ is the scale factor as a function of the conformal frame, i.e. we have $C(\eta) = a^2(t)$. In view of eqs.~(\ref{velconstraint}) and (\ref{conftimemetric}), the vector field (\ref{covariant}) appears to satisfy the constraint
\be \label{constraint}
	A_\mu \,A_\nu \,g^{\mu\nu} = - \frac{1}{\alp} ~,
\ee
since $g^{\mu\nu} = C^{-1}(\eta) \,\eta^{\mu\nu}$ for the FLRW metric in conformal time frame. Since the left-hand-side of (\ref{constraint}) is coordinate-frame independent, the constraint (\ref{constraint}) also characterises the FLRW time frame (\ref{flrwmetric}).

Note that in conformal time $\eta$ coordinates, the field strength (\ref{vectorfield}) becomes
\be \label{ftct}
	F_{0i} = - \frac{2}{\sqrt{\alp}} \;a^\prime \,u_i ~,
\ee
where the prime now means derivation with respect to $\eta$.

Thus, the cosmological background appears to break (spontaneously) the stringy gauge symmetry, leading to a massive vector field. This has been taken into account in the analysis of ref.~\cite{msy} as well as in this work, whenever the global (cosmological) background (\ref{covariant}) is used, like, for instance, in the section~\ref{sec:inflation} where we shall consider its role in an inflationary era of this string/fluctuating-D-particle universe.

However, when one considers local regions of space-time, such as a galaxy, of relevance to phenomenological tests of the model via lensing analyses, discussed in section~\ref{sec:lensing}, the space-time background is assumed static and spherically symmetric to a good approximation, of the form
\begin{align} \label{localmetric}
	g_{\alpha\beta} \,{\rm d}x^\alpha {\rm d}x^\beta &= - e^{\nu(r)} {\rm d}t^2 + e^{\zeta(r)} a^2(t) \,({\rm d}r^2 + r^2 {\rm d}\theta^2 + r^2 \sin^2 \theta {\rm d}\varphi^2) \nn
	&= - e^{\nu\left(\sqrt{x^2+y^2+z^2}\right)} {\rm d}t^2 + e^{\zeta\left(\sqrt{x^2+y^2+z^2}\right)} \,a^2(t) ({\rm d}x^2 + {\rm d}y^2 + {\rm d}z^2) ~,
\end{align}
where we kept track of the (small) universe expansion at the galactic era through the dependence of the metric on the scale factor $a(t)$; here $r = r(x,y,z) = \sqrt{x^2+y^2+z^2}$ in terms of Cartesian coordinates. The reason why we expressed the metric in Cartesian coordinates will become clear later. This metric is used in section~\ref{sec:lensing}.

In such metrics, the (recoil) fluctuations of the D-particle due to their interactions with open strings, representing galactic matter in our brane world, correspond to world-sheet deformations of gauge fields that can be well approximated by\footnote{We ignore, as subleading, any term in the geodesics of the D-particle associated with the local acceleration induced by the galactic mass.}
\be \label{ai_gen1}
	A_i (\vec x, t) \simeq \frac{1}{\alp} \,g_{ij} (\vec x, t) \,Y^j(t) \Theta (t - t_{\rm c}) \Big|_{\propto u_i} = \frac{1}{\alp} \,g_{ij} (\vec x, t) \,u^j \left( t \,\frac{a(t_c)^2}{a(t)^2} - t_{\rm c} \right) ~, \quad t > t_{\rm c} ~,
\ee
with $u^i$ ($i =x, y, z$) the spatial components of the D-particle recoil $3$-velocity in Cartesian coordinates. In constructing the local velocity field above, we took into account the non clustering effects of the $D$-particles, by maintaining their spatial trajectories as given by the geodesics $Y^i(t)$ in the global case (\ref{geodesics}), but replacing the FLRW metric by the local metric (\ref{localmetric}). However, for populations of D-particles in the neighbourhood of a galaxy, which are relevant for the lensing phenomenology of the D-material universe we are interested in here, the impact time $t_{\rm c} $ is of the same order of magnitude as the FLRW cosmic time of a galaxy of given redshift $z$~\cite{timered}
\begin{align} \label{timeredshift}
	t_c \sim \frac{2}{H_0 \,[1+(1+z)^2]}
\end{align}
where $H_0$ is the present value of the Hubble constant. In what follows, we discuss redshifts $z \ll 10$ so essentially this amounts to setting $a(t_{\rm c}) \sim a_0 =1$ in an order of magnitude, where $t_0$ is the present time, since in galactic eras the expansion of the universe is assumed small. Because of this, we kept the explicit (cosmic) time dependence of $Y^i(t)$ in eq.~(\ref{ai_gen1}), in contrast to the global case (\ref{geodesics}). In lensing analysis, discussed in section~\ref{sec:lensing}, the cosmic time $t$ appearing in eq.~(\ref{ai_gen1}) and subsequent formulae, coincides with the time of observation, that is the present time $t=t_0$.

The vector field $A_\mu$ in (\ref{ai_gen1}) also satisfies the constraint (\ref{constraint}). It can then be shown after detailed computations that, on account of the constraint, any terms $\partial_i A_t$ in $F_{ti}$ are subleading compared to $\partial_t A_i$, and thus from (\ref{ai_gen1}), (\ref{localmetric}) we conclude that to a very good approximation
\be \label{foicorrect1}
	F_{ti}(\vec x, t) \sim \frac{1}{\alp} \,g_{ij} (\vec x, t) \,u^j \left( \frac{a^2(t_c)}{a^2(t)} - 2 H(t) \,t_{\rm c} \right) ~,
\ee
where $H(t) \equiv \frac{\dot a(t)}{a(t)}$ is the Hubble parameter at (cosmic) time $t$. As already mentioned, for lensing measurements the time $t$ is the time of observation, that is today $t=t_0$, for which $a(t_0)=1$ in our normalisation. Thus, using eq.~(\ref{timeredshift}) and the fact that for an expanding universe $a(t_{\rm c}) = a_0 \frac{1}{1 + z}$, with $z$ the redshift of the galaxy in the neighborhood of which we consider local populations of D-particles, we obtain from (\ref{foicorrect1})
\be \label{foicorrect}
	F_{ti} (\vec x, t) \sim \frac{1}{\alp} \,g_{ij} (\vec x, t) \,u^j \left[ \frac{1 - 3(1 + z)^2}{(1 + z)^2 \,( 1 + (1 + z)^2)} \right] ~.
\ee
In a similar manner, the ``magnetic type'' field strength components $F_{ij}$ are much smaller than $F_{ti}$ as becomes clear from the expression
\begin{align} \label{magnetic}
	F_{ij} &= \partial_i A_j - \partial_j A_i = \frac{1}{\alp} \,a^2(t) \left[ t\,\frac{a^2(t_{\rm c})}{a^2(t)} - t_{\rm c} \right] \partial_{[i} \Big( e^{\zeta(r)} \Big) u_{j]} \nn
	&= a^2(t) \,\frac{e^{\zeta(r)}}{\alp} \left( u_i x^j - u_j x^i \right) \frac{\zeta'(r)}{r} \,\left[ t \,\frac{a^2(t_c)}{a^2(t)} - t_{\rm c} \right]~,
\end{align}
where $i,j,k,m,n$ denote Cartesian spatial 3-coordinates, the prime in $\zeta^\prime (r)$ denotes derivative with respect to $r$ and $[i...j]$ denotes antisymmetrisation in the respective indices. We now notice that $\zeta^\prime (r) \propto -\frac{\mathcal M}{r^2}$, ${\mathcal M}$ being the mass of the galaxy, is the gravitational acceleration induced by the galaxy on a D-particle, which is negligible compared to the terms we keep here, thus implying (\ref{magnetic}) leads to suppressed contributions in the dynamics as compared to the $F_{0i}$ terms, which we concentrate upon from now on.

For late (galaxy formation) eras of the universe, we consider populations of D-particles with fluctuating recoil velocities, which are assumed to be Gaussian stochastic for simplicity
\begin{equation} \label{stat}
	\llang u^m u^n \rrang = \sigma_0^2 (t) \,\delta^{mn}~, \qquad \llang u^m \rrang = 0~, \qquad \sigma_0^2 (t) = a(t)^{-3} \,|\beta|~,
\end{equation}
in order to macroscopically maintain Lorentz invariance in populations of D-particle defects. Notice that here, $u^i$ are Cartesian coordinates, which is the reason we previously were interested in expressing the local metric fluctuations in terms of such coordinates. The transformation of the result to spherical polar coordinates is of course straightforward.

The reader is invited at this stage to notice the presence of the (inverse cube of the) scale factor in eq.~(\ref{stat}), which will play a r\^ole later, which arises from the fact that the statistical fluctuations are proportional to the cosmic density of defects at a global scale~\cite{msy}, with $a(t)^3$ denoting the proper volume in a FLRW universe. In a semi-microscopic treatment, this scaling of $\sigma_0$ can be justified by noting that essentially
\be \label{avvel}
	\llang u_i u_j g^{ij} \rrang (t) \sim V_{\mathcal D}^{-1} \,\int _{\mathcal D} \,{\mathcal P} \,\overline u_i \overline u_j \,g^{ij} ~,
\ee
where $g_{ij} (\vec x, t)$ is the metric (\ref{localmetric}), ${\mathcal D}$ is a \emph{spatial} domain (with (proper) three-volume $V_{\mathcal D}$) upon which the (statistical) average over D-particle populations is considered at any given moment in cosmic time $t$, and $\mathcal P = \frac{n_D}{n_\gamma}$ is a probability of recoil of a D-particle under its interaction with low-energy cosmic photons, assumed to be the main contribution for the generation of the recoil field, with $n_D$, $n_\gamma$ the corresponding densities of D-particles and photons respectively. The quantity $\overline u_i$ is the spatial recoil velocity arising from a single scattering event of a photon with a D-particle, as described in refs.~\cite{kogan,gravanis}. It is proportional to the momentum $\overline p_i$ transfer during the scattering
\be \label{trnsf}
	\overline u_i = \frac{\Delta \overline p_i}{M_{\rm s}} \,g_{{\rm s}0} = \frac{{\widetilde \xi}_0 \,\overline p_i}{M_{\rm s} }\,g_{{\rm s}0} = \frac{{\widetilde \xi}_0 \,\overline p_i^{\rm phys}}{a(t) \,M_{\rm s}} \,g_{{\rm s}0}
\ee
with ${\widetilde \xi }_0 < 1$ is a space-time local constant ``fudge'' factor (hence independent of the universe's expansion), characteristic of the microscopic theory, and $p_i^{\rm phys}$ denotes the ``physical'' momentum observed by a cosmological observer who is comoving with the Hubble flow for an expanding universe with scale factor $a(t)$. The quantity $M_{\rm s} / g_{{\rm s} 0}$ is the mass of a D-particle, with $M_{\rm s} = 1 / \sqrt{\alp}$ the string scale and $g_{{\rm s}0} < 1$ the string coupling, assumed weak so that string-loop perturbation theory (and thus world-sheet formalism of recoil~\cite{kogan}) is valid. From eqs.~(\ref{localmetric}), (\ref{avvel}) and (\ref{trnsf}), and taking into account the scaling with the scale factor $a(t)$ of the densities of the (non interacting among themselves) D-particles, $n_D = n_D^{(0)} a^{-3}(t)$, and the photons (radiation) $n_\gamma = n_\gamma^{(0)} \,a^{-4}(t)$, with the superscript $(0)$ denoting present-day quantities, we obtain
\be \label{u2}
	\llang u_i u_j g^{ij} \rrang (t) \,\sim \frac{1}{a^3(t)} \,\frac{n_D^{(0)}}{n_\gamma^{(0)}} \frac{{\widetilde \xi}_0^2 \,| \overline{p}_i^{\rm phys} |^2 }{M_{\rm s} ^2}\,g_{{\rm s}0}^2 ~,
\ee
with $|p_i^{\rm phys} |^2 $ the square of the amplitude of the physical spatial momenta, computed with the time-independent part of the spatial metric (\ref{localmetric}) with $a^2(t)$ factored out. Comparing (\ref{u2}) with (\ref{stat}) we obtain an estimate for $|\beta|$ at late (galactic) epochs of the universe
\be \label{betagala}
	|\beta| \sim \frac{1}{3} \,\frac{n_D^{(0)}}{n_\gamma^{(0)}} \frac{{\widetilde \xi}_0^2 \,|\overline{p}_i^{\rm phys}|^2}{M_{\rm s}^2 } \,g_{{\rm s}0} ^2 ~.
\ee
We remind the reader that eq.~(\ref{betagala}) relies on the assumption that the dominant contributions to the recoil velocity field and its statistical fluctuations come from scatterings of D-particles with the abundant background of cosmic photons, taken for concreteness to be mostly CMB for the galactic era. In this case $|\overline{p}_i^{\rm phys}| $ denotes an average energy $\overline E$ of such photons as observed in the present day, i.e. $\overline E ({\rm eV}) = 1.24 / \lambda (\mu {\rm m}) $, with $\lambda (\mu {\rm m})$ a typical wavelength of the CMB photon, $\lambda \sim 1.9 \,{\rm mm}$. This yields
\be \label{cmbmom}
	\overline E^{\rm CMB} = \sqrt{|{p_i^{\rm phys}}^{\rm CMB}|^2} \sim 7 \times 10^{-4}~{\rm eV} ~.
\ee

We now remark that the $a(t)^{-3}$ dependence of $\sigma_0$ in (\ref{stat}) is over and above any inhomogeneities that may characterise local populations of D-particles in the neighbourhood of galaxies we shall concentrate upon when performing the lensing analysis. The latter may be incorporated in a mild dependence of $\sigma_0$ from galaxy to galaxy, which, as we show in section~\ref{sec:lensing}, may arise from uncertainties in cosmological measurements. It is also understood that the statistical averages above, leading to eqs.~(\ref{stat}), (\ref{betagala}) and (\ref{cmbmom}) are only applied at the end of the pertinent computations.

At this point we should make the following remark. The quantum fluctuations about such averaged quantities over populations of D-particles can be described in terms of the low-energy string effective action, cf. eq.~(\ref{action1}) discussed in the main text. The nonlinear Born-Infeld dynamics encoded in this action might then be responsible~\cite{odintsov} for the formation of \emph{quantum condensates} of the recoil velocity field that characterise the early universe epoch, which are distinct from the statistical averages (\ref{stat}) that correspond to the classical part of a condensate $\llang F_{\mu\nu} \,F^{\mu\nu} \rrang$ of the D-particle recoil velocity field. In our considerations in this article, the inflationary epoch is characterised by very dense populations of D-particles, and as such one may consider the classical, statistical effects as dominant.

%%%%%%%%%%%%%%%%%%%%%%%%%%%%%%%%%%%%%%%%%%%%%%%%%%%%%%%%%%
\section{Large-dilaton-induced Starobinsky-like inflation for small condensates \label{sec:dilaton}}

Although small condensate inflation induced by the D-particle population alone, with a zero dilaton, is not a viable scenario, as we have seen above, nevertheless one may~\cite{dflation} obtain Starobinsky-type inflation induced by the dilaton in a slow-rolling regime where the dilaton assumed large negative values. In this case, a crucial r\^ole is played by the D-particle small condensates in assisting this inflation in the sense of providing the means for a potential minimum towards which the dilaton field (which plays here the r\^ole of an inflaton) rolls slowly. The details have been presented in ref.~\cite{dflation} and will not be repeated here, however, for completeness, we shall outline below the main features.

In the case of a non trivial dilaton $\phi$, a convenient starting point is the $\sigma$-model-frame effective action (\ref{action1}), expanded (for small condensate fields) up to quadratic order in $F^2$ recoil field strength
\begin{align} \label{action1b}
	S_{\rm eff~4dim} \simeq \int d^4 x \,\sqrt{-g} &\left[ - \frac{T_3 \,e^{-\phi}}{g_{{\rm s}0}} - \frac{{\tilde \Lambda} \,e^{-2 \phi}}{\kappa^2_0} - \mathcal D - \mathcal A_s + \left( \frac{\alpha \,T_3 \,e^{-\phi}}{g_{{\rm s}0}} + \frac{e^{-2 \phi}}{\kappa^2_0} \right) R(g) \right. \nonumber \\
	& - \left. \frac{T_3 \,e^{-\phi}}{g_{{\rm s}0}} \frac{\left( 2\pi \alp \right)^2 F^{\mu\nu} F_{\mu\nu}}{4} \left[ 1 - \alpha R(g) \right] + \mathcal{O}\left((\partial \phi)^2\right) \right]~,
\end{align}
where $\mathcal D$ are dilaton independent flux condensates in the brane, defined previously (cf. (\ref{fluxcond})), and
\be \label{massD}
	{\mathcal A}_{\rm s} \equiv \frac{M_{\rm s}}{g_{{\rm s}0}} \,e^{-\phi} \,n_{\rm s} ~,
\ee
with $n_s$ the proper space density of D-particles in the string frame, is a contribution to the brane vacuum energy due to the (rest) mass of the D-particles (cf. eq.~(\ref{dmassdensity})).

Upon considering vacuum condensates in a Hartree-Fock approximation, i.e. replacing $F_{\mu\nu}\/ F^{\mu\nu}$ by the condensate field in the presence of a dilaton $\llang F_{\mu\nu} F^{\mu\nu} \rrang \sim {\cal C}_{M4}(t)$, such that one can define the dimensionless condensate field $\sigma (t,x)$
\begin{align} \label{condfield}
	\sigma(t,x) \equiv \frac{1}{4} \,\alpha \kappa_{\rm eff}^2 \left( e^{-2\phi} {\cal J} \right) {\cal C}_{M4}(t)~,
\end{align}
where one recalls the definition of ${\cal J}$ as in eq.~(\ref{defJ}), the effective ation (\ref{action1b}) becomes~\cite{dflation}
\begin{align} \label{effqac}
	S_{\rm eff~4dim} \simeq \int d^4x \,\sqrt{-g} \;\frac{e^{-2 \phi}}{2 \kappa^2_{\rm eff}} \Big[ \left( 1 + 2 \sigma(x^\mu) \right) R - \frac{2}{\alpha} \,\sigma(x^\mu) - 2 \kappa_{\rm eff}^2 \left( 2 \bar{\mathcal{B}} + e^{2 \phi} {\cal D} \right) \,\Big] + \dots
\end{align}
where $\alpha = \sfrac{\pi^2}{6} \;\alp$ as before, the $\dots$ denote dilaton derivatives, which are subleading terms in the slow-roll inflationary phase we are interested in here, and where we have defined $\frac{1}{2 \kappa^2_{\rm eff}} \equiv \left( \frac{\alpha \,T_3 e^{\phi}}{g_{{\rm s}0}} + \frac{1}{\kappa^2_0} \right) \simeq \frac{1}{2} \,M_{\rm Pl}^{2}$ along with
\begin{equation} \label{cosmoconst}
	2{\bar {\mathcal B}} \equiv \frac{T_3 \,e^{\phi}}{g_{{\rm s}0}} - \frac{|{\tilde \Lambda}|}{\kappa^2_0} + e^{2\phi} \mathcal{A}_{\rm s} ~,
\end{equation}
which is an effective vacuum energy and is almost a cosmological constant for slowly rolling $\phi (t) \simeq \phi_0$ large and negative. The reader should recall at this stage that the dilaton equation imposes the condition (\ref{t3ltilde}), which allows us to express the parameter $|\tilde \Lambda|$ in terms of the brane tension $T_3 >0$.

We next pass into the Einstein frame, denoted by a supersctipt $E$, by redefining the metric~\cite{dflation}
\begin{align} \label{confmetric}
	g_{\mu\nu} \rightarrow g^{\rm E}_{\mu\nu} &= \left( 1 + 2 {\sigma (t,x)} \right) \,e^{-2 \phi_0} \,g_{\mu\nu} ~,
\end{align}
in which case the field $\sigma(t,x)$ becomes a dynamical scalar degree of freedom. We define a canonically-normalized scalar field $\varphi(t,x)$
\begin{align}
	\varphi (t,x) &\equiv \sqrt{\frac{3}{2}} \,{\rm ln} \left(1 + 2 \sigma (t,x) \right) ~,
\end{align}
so that the action (\ref{effqac}) becomes
\begin{align} \label{steps}
	S^{\rm E}_{\rm eff~4dim} = \frac{1}{2 \kappa_{\rm eff}^2} \,\int d^4x \sqrt{-g^{\rm E}} \left[ R^{\rm E} + g^{{\rm E}\,\mu\nu} \partial_\mu \varphi \,\partial_\nu \varphi - V \left( \varphi \right) \right] \nn
	+ {\mathcal O} \Big( g^{{\rm E} \,\mu\nu} \partial_\mu \varphi \,\partial_\nu \phi_0, \,g^{{\rm E} \,\mu\nu} \partial_\mu \phi_0 \,\partial_\nu \phi_0 \Big) ~,
\end{align}
with the effective potential $V(\varphi)$ in the inflationary regime of \emph{large negative values} of $\phi_0$ given approximately by~\cite{dflation}
\begin{align} \label{staropotent}
	V(\varphi ) \simeq \left[ \frac{1}{\alpha} \left( e^{\sqrt{\frac{2}{3}}\varphi} - 1 \right) + 4 \kappa_{\rm eff}^2 \bar {\mathcal B}^{\rm E} \right] e^{2 \phi_0} \,e^{- \sqrt{\frac{8}{3}} \varphi} + 2 \kappa_{\rm eff}^2 \,\mathcal D~,
\end{align}
where
\begin{align}
	2 \bar {\mathcal B}^{\rm E} \simeq - \frac{T_3 \,e^{\phi_0}}{2 \,g_{{\rm s}0}} + e^{2 \phi_0}\,\mathcal{A}^{\rm E} ~,
\end{align}
with ${\mathcal A}^{\rm E}$ the Einstein frame (\ref{confmetric}) (dilaton-independent) vacuum energy corresponding to the $\sigma$-model-frame quantity $\mathcal A_s$ (\ref{massD}) (cf. eq.~(\ref{dmassdensity2}))
\be \label{einsteinendens}
	{\mathcal A}^{\rm E} \sim \frac{M_{\rm s}}{g_{{\rm s}0}} \,e^{-\phi_0} \,{n_D^\star} M_{\rm Pl}^3
\ee
where $n_D^\star $ is the number density of D-particles per (reduced) Planck three-volume on the brane world (assumed more or less constant during inflation~\cite{dflation}).

It is important to note that, in order to arrive at eq.~(\ref{staropotent}), we took into account the conformal nature of the flux condensate term in the four-dimensional space-time action $\int d^4 x \sqrt{g} \,\mathcal D$ (under rescalings of the form (\ref{confmetric})), and have ignored terms that are more than quadratic in the vector potential. Moreover, as already emphasized, for our purposes here we concentrate on the slow-roll phase of the dilaton field $\phi_0$, so any potential-like terms with dilaton time-derivative factors are ignored. In this approximation we need not to worry about the cross-kinetic-terms $\partial_\mu \phi_0 \partial^\mu \varphi$, which can in any case be eliminated by a further redefinition (mixing) of the fields $\varphi$ and $\phi_0$~\cite{dflation}.

From the discussion in ref.~\cite{dflation}, an extra factor $\sqrt{2}$ needs to be absorbed into the dilaton normalisation in order to obtain a canonical kinetic term for this field, yielding finally
\begin{align} \label{staropotent3}
	\frac{1}{2 \kappa_{\rm eff}^2} \,V(\varphi ) &= \frac{1}{\alpha \,2 \kappa_{\rm eff}^2} \left( e^{\sqrt{\frac{2}{3}} \varphi} - 1 \right) e^{- \sqrt{\frac{8}{3}} \varphi} \,e^{\sqrt{2} \phi_0} \nn
	& \quad - \frac{T_3}{2 g_{{\rm s}0}} \,e^{- \sqrt{\frac{8}{3}} \varphi} \,e^{\frac{3}{\sqrt{2}} \phi_0} + {\mathcal A}^{\rm E} \,e^{- \sqrt{\frac{8}{3}} \varphi} \,e^{2 \sqrt{2} \phi_0} + {\mathcal D}~.
\end{align}
For weak condensates $\varphi \ll 1$, where the approximations in this article hold, and large negative values of the dilaton $\phi_0$, the reader will recognize in (\ref{staropotent3}) the Starobinsky-like form of the effective potential for the dilaton-driven inflation provided ${\mathcal A} + {\mathcal D} > 0$. This can fit the Planck data~\cite{Planck} due to the very small value of the tensor-to-scalar ratio $r$ predicted by this class of theories.

By minimizing the effective potential (\ref{staropotent3}) with respect to the condensate field, for fixed large negative values of the dilaton $\phi_0$, we observe that the minimum occurs for $\varphi \simeq 0$ (as required for consistency) provided that
\be \label{estimate}
	{\mathcal A}^{\rm E} \simeq \frac{T_3}{2 g_{{\rm s}0}} \left[ \frac{3 g_{{\rm s}0}}{\pi^2} \frac{M_{\rm s}^2 \,M_{\rm Pl}^2}{T_3} \,e^{- \frac{1}{\sqrt{2}} \phi_0} + 1 \right] e^{- \frac{1}{\sqrt{2}} \phi_0} ~.
\ee
Taking into account (\ref{einsteinendens}) with the rescaled dilaton $\phi_0 \to \sfrac{\phi_0}{\sqrt{2}}$, we have from (\ref{estimate})
\be \label{density}
	n_{\rm D}^\star \sim \,\frac{T_3}{2 \,M_{\rm s} \,M_{\rm Pl}^3} \left[ \frac{3 g_{{\rm s}0}}{\pi^2} \frac{M_{\rm s}^2 \,M_{\rm Pl}^2}{T_3} \,e^{- \frac{1}{\sqrt{2}} \phi_0} + 1 \right]
\ee
for the densities of D-particles on the brane world during the dilaton-driven inflation area.

If we adopt the standard relation (used in ref.~\cite{msy}) $(2 \pi \alp)^2 \,T_3 = 1$, that is $4 \pi^2 \,T_3 = M_{\rm s}^4$, which is consistent with weak condensate fields $\varphi$, as we have seen previously, then one obtains
\be
	n_{\rm D}^\star \sim \,\frac{1}{8 \pi^2} \left( \frac{M_{\rm s}}{M_{\rm Pl}} \right)^3 \left[ 12 \,g_{{\rm s}0} \left( \frac{M_{\rm Pl}}{M_{\rm s}} \right)^2 e^{- \frac{1}{\sqrt{2}} \phi_0} + 1 \right] ~.
\ee
Now for $M_{\rm s} \sim 10^{16} {\rm GeV} \,\gg H_I \sim 10^{13} ~{\rm GeV}$, $g_{{\rm s}0} \sim 0.1$ and nominal values of the dilaton field in the range $|\phi_0| \in [1, 10]$, we obtain $n_{\rm D}^\star \in [10^{-4}, 10^{-1}]$. Higher densities can be obtained for larger brane tensions $T_3$. Thus, we observe that, even for the weak recoil-velocity condensate fields case, during slowly-rolling-dilaton-driven inflation, the density of D-particle defects must be much higher than then corresponding one in the galactic era. This is a rather generic feature of the D-material universe.

%%%%%%%%%%%%%%%%%%%%%%%%%%%%%%%%%%%%%%%%%%%%%%%%%%%%%%%%%%%%%%%%%%


\begin{thebibliography}{99}

\bibitem{Planck} P.~A.~R.~Ade {\it et al.} [Planck Collaboration],
%``Planck 2015 results. XIII. Cosmological parameters,''
 arXiv:1502.01589 [astro-ph.CO].
%%CITATION = ARXIV:1502.01589;%%
%867 citations counted in INSPIRE as Planck_inflof 07 Nov 2015

\bibitem{bao} For a sample of characteristic references see: H.~J.~Seo
 and D.~J.~Eisenstein,
%``Probing dark energy with baryonic acoustic oscillations from future
%large galaxy redshift surveys,''
Astrophys.\ J.\ {\bf 598}, 720 (2003)
[astro-ph/0307460];
%%CITATION = ASTRO-PH/0307460;%%
%459 citations counted in INSPIRE as of 07 Nov 2015
L.~Anderson {\it et al.},
%``The clustering of galaxies in the SDSS-III Baryon Oscillation
%Spectroscopic Survey: Baryon Acoustic Oscillations in the Data
%Release 9 Spectroscopic Galaxy Sample,''
Mon.\ Not.\ Roy.\ Astron.\ Soc.\ {\bf 427}, no. 4, 3435 (2013)
[arXiv:1203.6594 [astro-ph.CO]];
%%CITATION = ARXIV:1203.6594;%%
%456 citations counted in INSPIRE as of 07 Nov 2015
W.~J.~Percival {\it et al.} [SDSS Collaboration],
%``Baryon Acoustic Oscillations in the Sloan Digital Sky Survey Data
%Release 7 Galaxy Sample,''
Mon.\ Not.\ Roy.\ Astron.\ Soc.\ {\bf 401}, 2148 (2010)
[arXiv:0907.1660 [astro-ph.CO]];
%%CITATION = ARXIV:0907.1660;%%
%1039 citations counted in INSPIRE as of 07 Nov 2015
W.~J.~Percival, S.~Cole, D.~J.~Eisenstein, R.~C.~Nichol,
J.~A.~Peacock, A.~C.~Pope and A.~S.~Szalay,
%``Measuring the Baryon Acoustic Oscillation scale using the SDSS and 2dFGRS,''
Mon.\ Not.\ Roy.\ Astron.\ Soc.\ {\bf 381}, 1053 (2007)
[arXiv:0705.3323 [astro-ph]], and references therein.
%%CITATION = ARXIV:0705.3323;%%
%551 citations counted in INSPIRE as of 07 Nov 2015

\bibitem{sna} For a sample of characteristic references see:
B.~P.~Schmidt {\it et al.} [Supernova Search Team Collaboration],
%``The High Z supernova search: Measuring cosmic deceleration and
%global curvature of the universe using type Ia supernovae,''
Astrophys.\ J.\ {\bf 507}, 46 (1998)
[astro-ph/9805200];
%%CITATION = ASTRO-PH/9805200;%%
%861 citations counted in INSPIRE as of 07 Nov 2015
S.~Perlmutter {\it et al.} [Supernova Cosmology Project
 Collaboration],
%``Measurements of Omega and Lambda from 42 high redshift supernovae,''
Astrophys.\ J.\ {\bf 517}, 565 (1999)
[astro-ph/9812133].
%%CITATION = ASTRO-PH/9812133;%%
%8815 citations counted in INSPIRE as of 07 Nov 2015
A.~G.~Riess {\it et al.} [Supernova Search Team Collaboration],
%``Type Ia supernova discoveries at z > 1 from the Hubble Space
%Telescope: Evidence for past deceleration and constraints on dark
%energy evolution,''
Astrophys.\ J.\ {\bf 607}, 665 (2004)
[astro-ph/0402512].
%%CITATION = ASTRO-PH/0402512;%%
%2782 citations counted in INSPIRE as of 07 Nov 2015
R.~A.~Knop {\it et al.} [Supernova Cosmology Project Collaboration],
%``New constraints on Omega(M), Omega(lambda), and w from an independent set of eleven high-redshift supernovae observed with HST,''
Astrophys.\ J.\ {\bf 598}, 102 (2003)
[astro-ph/0309368];
%%CITATION = ASTRO-PH/0309368;%%
%1209 citations counted in INSPIRE as of 07 Nov 2015 
M.~Hicken, W.~M.~Wood-Vasey, S.~Blondin, P.~Challis, S.~Jha, P.~L.~Kelly, A.~Rest and R.~P.~Kirshner,
%``Improved Dark Energy Constraints from ~100 New CfA Supernova Type Ia Light Curves,''
Astrophys.\ J.\ {\bf 700}, 1097 (2009) [arXiv:0901.4804
 [astro-ph.CO]], and references therein.
%%CITATION = ARXIV:0901.4804;%%
%544 citations counted in INSPIRE as of 07 Nov 2015

%\cite{Ade:2015lrj}
\bibitem{Planck_infl}
P.~A.~R.~Ade {\it et al.} [Planck Collaboration],
%``Planck 2015 results. XX. Constraints on inflation,''
arXiv:1502.02114 [astro-ph.CO].
%%CITATION = ARXIV:1502.02114;%%
%82 citations counted in INSPIRE as of 16 Apr 2015

\bibitem{mond} M.~Milgrom,
%``A Modification of the Newtonian dynamics as a possible alternative
%to the hidden mass hypothesis,''
Astrophys.\ J.\ {\bf 270}, 365 (1983);
%%CITATION = ASJOA,270,365;%%
%1262 citations counted in INSPIRE as of 07 Nov 2015
%``A Modification of the Newtonian dynamics: Implications for galaxies,''
Astrophys.\ J.\ {\bf 270}, 371 (1983).
%%CITATION = ASJOA,270,371;%%
%427 citations counted in INSPIRE as of 07 Nov 2015

\bibitem{teves} J.~D.~Bekenstein,
%``Relativistic gravitation theory for the MOND paradigm,''
Phys.\ Rev.\ D {\bf 70}, 083509 (2004)
[Phys.\ Rev.\ D {\bf 71}, 069901 (2005)]
[astro-ph/0403694].
%%CITATION = ASTRO-PH/0403694;%%
%810 citations counted in INSPIRE as of 07 Nov 2015

\bibitem{famay} B.~Famaey and S.~McGaugh,
%``Modified Newtonian Dynamics (MOND): Observational Phenomenology and Relativistic Extensions,''
Living Rev.\ Rel.\ {\bf 15}, 10 (2012)
[arXiv:1112.3960 [astro-ph.CO]] and references therein.
%%CITATION = ARXIV:1112.3960;%%
%122 citations counted in INSPIRE as of 07 Nov 2015

\bibitem{ferreras} I.~Ferreras, N.~E.~Mavromatos, M.~Sakellariadou and M.~F.~Yusaf,
%``Confronting MOND and TeVeS with strong gravitational lensing over galactic scales: an extended survey,''
Phys.\ Rev.\ D {\bf 86} (2012) 083507
[arXiv:1205.4880 [astro-ph.CO]];
%%CITATION = ARXIV:1205.4880;%%
%10 citations counted in INSPIRE as of 07 Nov 2015
I.~Ferreras, N.~E.~Mavromatos, M.~Sakellariadou and M.~F.~Yusaf,
%``Incompatibility of Rotation Curves with Gravitational Lensing for TeVeS,''
Phys.\ Rev.\ D {\bf 80}, 103506 (2009)
[arXiv:0907.1463 [astro-ph.GA]];
%%CITATION = ARXIV:0907.1463;%%
%28 citations counted in INSPIRE as of 07 Nov 2015
N.~E.~Mavromatos, M.~Sakellariadou and M.~F.~Yusaf,
 %``Can the relativistic field theory version of modified Newtonian dynamics avoid dark matter on galactic scales?,''
 Phys.\ Rev.\ D {\bf 79} (2009) 081301
% doi:10.1103/PhysRevD.79.081301
 [arXiv:0901.3932 [astro-ph.GA]];
 %%CITATION = doi:10.1103/PhysRevD.79.081301;%%
 %20 citations counted in INSPIRE as of 02 Dec 2015
I.~Ferreras, M.~Sakellariadou and M.~F.~Yusaf,
%``The necessity of dark matter in MOND within galactic scales,''
Phys.\ Rev.\ Lett.\ {\bf 100}, 031302 (2008)
[arXiv:0709.3189 [astro-ph]].
%%CITATION = ARXIV:0709.3189;%%
%31 citations counted in INSPIRE as of 07 Nov 2015

\bibitem{tevescosmo} C.~Skordis,
%``Generalizing tensor-vector-scalar cosmology,''
Phys.\ Rev.\ D {\bf 77}, 123502 (2008)
[arXiv:0801.1985 [astro-ph]];
%%CITATION = ARXIV:0801.1985;%%
%61 citations counted in INSPIRE as of 07 Nov 2015 ``Teves cosmology :
%covariant formalism for the background evolution and linear
%perturbation theory,''
Phys.\ Rev.\ D {\bf 74}, 103513 (2006)
[astro-ph/0511591].
%%CITATION = ASTRO-PH/0511591;%%
%68 citations counted in INSPIRE as of 07 Nov 2015

\bibitem{dodelson} S.~Dodelson and M.~Liguori,
%``Can Cosmic Structure form without Dark Matter?,''
Phys.\ Rev.\ Lett.\ {\bf 97}, 231301 (2006)
[astro-ph/0608602].
%%CITATION = ASTRO-PH/0608602;%%
%98 citations counted in INSPIRE as of 07 Nov 2015

\bibitem{ms} N.~Mavromatos and M.~Sakellariadou,
%``Relativistic Modified Newtonian Dynamics from String Theory?,''
Phys.\ Lett.\ B {\bf 652}, 97 (2007)
[hep-th/0703156 [HEP-TH]].
%%CITATION = HEP-TH/0703156;%%
%30 citations counted in INSPIRE as of 07 Nov 2015

\bibitem{dfoam} J.~R.~Ellis, N.~E.~Mavromatos and M.~Westmuckett,
%``A Supersymmetric D-brane model of space-time foam,''
Phys.\ Rev.\ D {\bf 70}, 044036 (2004)
[gr-qc/0405066];
%%CITATION = GR-QC/0405066;%%
%69 citations counted in INSPIRE as of 07 Nov 2015
%``Potentials between D-branes in a supersymmetric model of space-time foam,''
Phys.\ Rev.\ D {\bf 71}, 106006 (2005)
[gr-qc/0501060];
%%CITATION = GR-QC/0501060;%%
%50 citations counted in INSPIRE as of 07 Nov 2015
J.~R.~Ellis, N.~E.~Mavromatos, D.~V.~Nanopoulos and M.~Westmuckett,
%``Liouville cosmology at zero and finite temperatures,''
Int.\ J.\ Mod.\ Phys.\ A {\bf 21}, 1379 (2006)
[gr-qc/0508105].
%%CITATION = GR-QC/0508105;%%
%44 citations counted in INSPIRE as of 07 Nov 2015

%\cite{Mavromatos:2010nk}
\bibitem{mitsou} 
 N.~E.~Mavromatos, V.~A.~Mitsou, S.~Sarkar and A.~Vergou,
 %``Implications of a Stochastic Microscopic Finsler Cosmology,''
 Eur.\ Phys.\ J.\ C {\bf 72}, 1956 (2012)
 doi:10.1140/epjc/s10052-012-1956-7
 [arXiv:1012.4094 [hep-ph]].
 %%CITATION = doi:10.1140/epjc/s10052-012-1956-7;%%
 %35 citations counted in INSPIRE as of 01 Dec 2015

\bibitem{dmatter}
G.~Shiu and L.~T.~Wang,
 %``D matter,''
 Phys.\ Rev.\ D {\bf 69}, 126007 (2004)
 doi:10.1103/PhysRevD.69.126007
 [hep-ph/0311228].
 %%CITATION = doi:10.1103/PhysRevD.69.126007;%%
 %33 citations counted in INSPIRE as of 01 Dec 2015

%\cite{Mavromatos:2012ha}
\bibitem{msy} 
N.~E.~Mavromatos, M.~Sakellariadou and M.~F.~Yusaf,
%``Stringy Models of Modified Gravity: Space-time defects and
%Structure Formation,''
JCAP {\bf 1303}, 015 (2013)
[arXiv:1211.1726 [hep-th]].
%%CITATION = ARXIV:1211.1726;%%
%5 citations counted in INSPIRE as of 16 mar 2015

\bibitem{kogan} I.~I.~Kogan, N.~E.~Mavromatos and J.~F.~Wheater,
%``D-brane recoil and logarithmic operators,''
Phys.\ Lett.\ B {\bf 387}, 483 (1996)
[hep-th/9606102].
%%CITATION = HEP-TH/9606102;%%
%131 citations counted in INSPIRE as of 10 Oct 2015

%\cite{Gravanis:2002ii}
\bibitem{gravanis} 
E.~Gravanis and N.~E.~Mavromatos,
%``Higher-order logarithmic conformal algebras from Robertson-Walker sigma-model backgrounds,''
JHEP {\bf 0206}, 019 (2002).
%%CITATION = JHEPA,0206,019;%%
%11 citations counted in INSPIRE as of 14 Jun 2015

%\cite{Ellis:2014kha}
\bibitem{dflation} 
J.~Ellis, N.~E.~Mavromatos and D.~V.~Nanopoulos,
%``Starobinsky-Like Inflation in Dilaton-Brane Cosmology,''
Phys.\ Lett.\ B {\bf 732}, 380 (2014)
[arXiv:1402.5075 [hep-th]];
%%CITATION = ARXIV:1402.5075;%%
%17 citations counted in INSPIRE as of 26 Oct 2015
%``D-Flation,''
JCAP {\bf 1411}, no. 11, 014 (2014)
[arXiv:1406.0487 [hep-th]].
%%CITATION = ARXIV:1406.0487;%%
%1 citations counted in INSPIRE as of 22 Jun 2015 

\bibitem{tseytlin} See, for instance: A.~A.~Tseytlin,
%``Dirac-Born-Infeld action, supersymmetry and string theory,''
arXiv:hep-th/9908105, in \emph{Yuri Golfand memorial volume},
ed. M. Shifman, World Scientific, 2000; also in \emph{The many
faces of the superworld} (Shifman, M.A. (ed.): 417-452), and
references therein.
%%CITATION = HEP-TH/9908105;%%

\bibitem{Cheung}
Y.~K.~Cheung, M.~Laidlaw and K.~Savvidy,
%``Open string gravity?,''
JHEP {\bf 0412}, 028 (2004)
[arXiv:hep-th/0406245].

\bibitem{refractive} J.~R.~Ellis, N.~E.~Mavromatos and D.~V.~Nanopoulos,
%``Derivation of a Vacuum Refractive Index in a Stringy Space-Time Foam Model,''
Phys.\ Lett.\ B {\bf 665}, 412 (2008)
[arXiv:0804.3566 [hep-th]];
%%CITATION = ARXIV:0804.3566;%%
%84 citations counted in INSPIRE as of 07 Nov 2015
T.~Li, N.~E.~Mavromatos, D.~V.~Nanopoulos and D.~Xie,
%``Time Delays of Strings in D-particle Backgrounds and Vacuum Refractive Indices,''
Phys.\ Lett.\ B {\bf 679}, 407 (2009)
[arXiv:0903.1303 [hep-th]].
%%CITATION = ARXIV:0903.1303;%%
%34 citations counted in INSPIRE as of 07 Nov 2015


\bibitem{timered} See, for instance: M.~Carmeli, J.~G.~Hartnett and F.~J.~Oliveira,
%``The Cosmic time in terms of the redshift,''
Found.\ Phys.\ Lett.\ {\bf 19}, 277 (2006)
[gr-qc/0506079] and references therein.
%%CITATION = GR-QC/0506079;%%
%5 citations counted in INSPIRE as of 12 juin 2015

\bibitem{NFW} J.\ F\ Navarro, C.\ S.\ Frenk, S.\ D.\ M.\ White, Astrophys.\
J.\ {\bf 493} (1996) 563.

\bibitem{sal} E.\ E.\ Salpeter, Astrophys.\ J.
{\bf 121}, 161 (1955).

\bibitem{chab03} G.\ Chabrier, Publ.\ Astron.\ Soc.\ Pac.\ {\bf 115}
(2003) 763.

\bibitem{ig} D.~Leier, I.~Ferreras, P.~Saha and E.~E.~Falco
Astrophys. J. 740 (2011) 97;
[arXiv:1102.3433 [astro-ph.GA]]

\bibitem{rus03} D.\ Rusin {\sl et. al.}, Astrophys.\ J.\ {\bf 587}
(2003) 143.
 
\bibitem{gh}%\cite{Gibbons:1977mu}
G.~W.~Gibbons and S.~W.~Hawking,
%``Cosmological Event Horizons, Thermodynamics, and Particle Creation,''
Phys.\ Rev.\ D {\bf 15}, 2738 (1977);
%%CITATION = PHRVA,D15,2738;%%
%1575 citations counted in INSPIRE as of 17 mar 2015
see also in a more general context of Euclidean quantum gravity: 
G.~W.~Gibbons and S.~W.~Hawking, \emph{Euclidean quantum gravity,} (Singapore, Singapore: World Scientific (1993)).
%3 citations counted in INSPIRE as of 17 mar 2015

\bibitem{hawking} S.~W.~Hawking,
%``Particle Creation by Black Holes,''
Commun.\ Math.\ Phys.\ {\bf 43}, 199 (1975)
[Commun.\ Math.\ Phys.\ {\bf 46}, 206 (1976)];
%%CITATION = CMPHA,43,199;%%
%5686 citations counted in INSPIRE as of 02 Nov 2015



%\cite{Komatsu:2010fb}
\bibitem{Komatsu:2010fb} 
E.~Komatsu {\it et al.} [WMAP Collaboration],
%``Seven-Year Wilkinson Microwave Anisotropy Probe (WMAP) Observations: Cosmological Interpretation,''
Astrophys.\ J.\ Suppl.\ {\bf 192}, 18 (2011)
doi:10.1088/0067-0049/192/2/18
[arXiv:1001.4538 [astro-ph.CO]].
%%CITATION = doi:10.1088/0067-0049/192/2/18;%%
%5312 citations counted in INSPIRE as of 20 Nov 2015

%\cite{Hinshaw:2012aka}
\bibitem{Hinshaw:2012aka} 
G.~Hinshaw {\it et al.} [WMAP Collaboration],
%``Nine-Year Wilkinson Microwave Anisotropy Probe (WMAP) Observations: Cosmological Parameter Results,''
Astrophys.\ J.\ Suppl.\ {\bf 208}, 19 (2013)
doi:10.1088/0067-0049/208/2/19
[arXiv:1212.5226 [astro-ph.CO]].
%%CITATION = doi:10.1088/0067-0049/208/2/19;%%
%1774 citations counted in INSPIRE as of 20 Nov 2015

%\cite{Elizalde:2003ku}
\bibitem{odintsov}
E.~Elizalde, J.~E.~Lidsey, S.~Nojiri and S.~D.~Odintsov,
%``Born-Infeld quantum condensate as dark energy in the universe,''
Phys.\ Lett.\ B {\bf 574}, 1 (2003)
[hep-th/0307177].
%%CITATION = HEP-TH/0307177;%%
%89 citations counted in INSPIRE as of 22 Jun 2015



\bibitem{acs} A.~Campbell-Smith and N.~E.~Mavromatos,
 %``D-brane recoil and supersymmetry breaking as a relaxation process,''
 Phys.\ Lett.\ B {\bf 476}, 149 (2000)
 doi:10.1016/S0370-2693(00)00127-1
 [hep-th/9908139].
 %%CITATION = doi:10.1016/S0370-2693(00)00127-1;%%
 %10 citations counted in INSPIRE as of 01 Dec 2015

\bibitem{globalmono} M.~Barriola and A.~Vilenkin,
 %``Gravitational Field of a Global Monopole,''
 Phys.\ Rev.\ Lett.\ {\bf 63}, 341 (1989).
 doi:10.1103/PhysRevLett.63.341
 %%CITATION = doi:10.1103/PhysRevLett.63.341;%%
 %434 citations counted in INSPIRE as of 01 Dec 2015


\bibitem{papavass} P.~O.~Mazur and J.~Papavassiliou,
 %``Gravitational scattering on a global monopole,''
 Phys.\ Rev.\ D {\bf 44}, 1317 (1991).
 [doi:10.1103/PhysRevD.44.1317];
 %%CITATION = doi:10.1103/PhysRevD.44.1317;%%
 %10 citations counted in INSPIRE as of 01 Dec 2015
 H.~Ren,
 %``Fermions in a global monopole background,''
 Phys.\ Lett.\ B {\bf 325}, 149 (1994)
 doi:10.1016/0370-2693(94)90085-X
 [hep-th/9312074];
 %%CITATION = doi:10.1016/0370-2693(94)90085-X;%%
 %5 citations counted in INSPIRE as of 01 Dec 2015
see also: A.~A.~Roderigues Sobreira and E.~R.~Bezerra de Mello,
 %``The Classical and quantum analysis of a charged particle on the space-time produced by a global monopole,''
 Grav.\ Cosmol.\ {\bf 5}, 177 (1999)
 [hep-th/9809212];
 %%CITATION = HEP-TH/9809212;%%
 %4 citations counted in INSPIRE as of 01 Dec 2015
E.~R.~Bezerra de Mello and C.~Furtado,
 %``The Nonrelativistic scattering problem by a global monopole,''
 Phys.\ Rev.\ D {\bf 56}, 1345 (1997).
 doi:10.1103/PhysRevD.56.1345
 %%CITATION = doi:10.1103/PhysRevD.56.1345;%%
 %23 citations counted in INSPIRE as of 01 Dec 2015

\bibitem{moedal} B.~Acharya {\it et al.} [MoEDAL Collaboration],
 %``The Physics Programme Of The MoEDAL Experiment At The LHC,''
 Int.\ J.\ Mod.\ Phys.\ A {\bf 29}, 1430050 (2014)
 doi:10.1142/S0217751X14300506
 [arXiv:1405.7662 [hep-ph]].
 %%CITATION = doi:10.1142/S0217751X14300506;%%
 %13 citations counted in INSPIRE as of 01 Dec 2015

\bibitem{dinmoedal} See talk by N.E.~Mavromatos, in MoEDAL
 Collaboration Meeting, CERN, Geneva, Switzerland, June 19-21 2015,
 available in: {\tt https://indico.cern.ch/event/320391/}.

\end{thebibliography}
\end{document}